\documentclass[preprint]{elsarticle}%
\usepackage[T1]{fontenc}
\usepackage[margin=4.2cm]{geometry}% by courtesy of Mico

\usepackage{listings}
\usepackage{amsmath,amssymb,amsfonts}
\usepackage{algorithmic}
\usepackage{graphicx}
\usepackage{textcomp}
\usepackage{booktabs}
\usepackage{threeparttable}
\usepackage{amsthm}
\usepackage{url}
\usepackage{makecell}
\usepackage[table]{xcolor}
\usepackage{pifont}
\usepackage{xspace}
\usepackage{float}
\usepackage{footmisc}
\usepackage{subcaption}
\usepackage[most]{tcolorbox}
\usepackage{siunitx} 
\usepackage{array}
\usepackage{hyperref}
\biboptions{sort&compress}
\newtheorem{definition}{Definition}[section]
\newcommand{\missingcell}{\cellcolor{gray!25}}
\newcolumntype{R}[1]{>{\raggedleft\arraybackslash}p{#1}}
\begin{document}
\begin{frontmatter}

\title{Entropy Collapse in Mobile Sensors: The Hidden Risks of Sensor-Based Security}

\author[1]{Carlton Shepherd\corref{cor1}\fnref{fn1}}
\ead{carlton.shepherd@ncl.ac.uk}
\author[1]{Elliot A.\ J.\ Hurley}
\ead{e.a.j.hurley2@ncl.ac.uk}
\address[1]{School of Computing, Newcastle University, Newcastle-upon-Tyne, UK}

% Optional: Include corresponding author and footnotes if needed.
\cortext[cor1]{Corresponding author}
\fntext[fn1]{ORCID: 0000-0002-7366-9034}

\begin{abstract}Mobile sensor data has been proposed for security-critical applications such as device pairing, proximity detection, and continuous authentication. However, the foundational premise that these signals provide sufficient entropy remains under-explored.  In this work, we systematically analyse the entropy of mobile sensor data using four datasets from multiple application contexts (UCI-HAR, SHL, Relay, and PerilZIS).  Using direct computation and estimation, we report entropy values---max, Shannon, collision, and min-entropy---for an exhaustive range of sensor combinations. We demonstrate that the entropy of mobile sensors remains far below what is considered secure by modern standards for security applications, even when many sensors are combined. In particular, we observe an alarming divergence between average-case Shannon entropy and worst-case min-entropy. Single-sensor min-entropy varies between 3.408--4.483 bits despite Shannon entropy being several multiples higher. We also show that redundancies between sensor modalities contribute to a $\approx$75\% reduction between Shannon and min-entropy.  Indeed, min-entropy plateaus between 8.1--23.9 bits when combining up to 22 modalities, while Shannon entropy can exceed 80 bits. Adding sensors typically increases Shannon entropy but moves min-entropy by only $\approx$1--2 bits per added modality, evidencing entropy collapse under redundancy. Our results reveal that adversaries may feasibly predict sensor signals through an exhaustive exploration of the measurement space. Our work also calls into question the widely held assumption that adding more sensors inherently yields higher security. Ultimately, we strongly urge caution when relying on mobile sensor data for security applications.
%This brings joint min-entropy well below 10 bits in many cases and, in the best case, yielding only $\approx$24 bits when combining 20+ sensor modalities.
\end{abstract}

\begin{keyword}
Entropy \sep Sensors \sep Mobile security
\end{keyword}

\end{frontmatter}
\section{Introduction}
\label{sec:intro}
Modern mobile devices expose rich motion and ambient signals from accelerometers, gyroscopes, magnetometers, and more. Beyond activity recognition and context inference, e.g.~\cite{gjoreski2018university,wang2019enabling,anguita2013public,yin2024systematic}, a long line of work proposes using these signals for security-critical tasks, such as cryptographic key generation, zero-interaction device pairing, and continuous authentication~\cite{gurulian2017effectiveness,gurulian2018good,mehrnezhad2015tap,mayrhofer2009shake,shrestha2014drone,shrestha2018sensor,shepherd2017applicability,shi2011senguard,miettinen2014conxsense,li2013unobservable,li2020t2pair,riva2012progressive,krhovjak2007sources, xu2021key,abuhamad2020sensor,mayrhofer2021adversary}. Traditional key generation protocols~\cite{mayrhofer2009shake,xu2021key} have historically relied on shared motion patterns in order to establish secure communication between co-located devices. Meanwhile, other proposals have incorporated environmental phenomena, such as magnetic field characteristics and thermal fluctuations, to mitigate relay attacks~\cite{shrestha2014drone,shrestha2018sensor,markantonakis2024using,gurulian2017effectiveness,shepherd2017applicability} and to reduce the frequency of user authentication prompts~\cite{krhovjak2007sources,riva2012progressive,shi2011senguard,miettinen2014conxsense,li2013unobservable}.  

Despite these advances, a crucial assumption remains under-examined: that sensor data is robust against adversaries aiming to forge or reconstruct such data through statistical inference.  Existing proposals rely heavily on heuristic assessments or model classification metrics, e.g.~\cite{markantonakis2024using,mehrnezhad2015tap,gurulian2018good,shrestha2014drone,truong2014comparing,abuhamad2020sensor,mayrhofer2021adversary}, which do not address two basic questions: (1) how much entropy do sensors actually provide and (2) to what extent do multimodal sensor combinations improve entropy? For the latter, multimodal sensing has been proposed as a means to counteract sensor-specific weaknesses~\cite{shrestha2014drone,markantonakis2024using,truong2014comparing,mehrnezhad2015tap}, but the quantitative security benefits of combining multiple sensors have not been rigorously evaluated. The security implications of low-entropy sensor data can be both direct and severe. If such data is used as authentication evidence, or to derive secret keys, an adversary may attempt to enumerate the entire space of possible sensor measurement values in order to recover authentication or key material. Modern standards provide important quantitative thresholds for what can be considered `secure' for such reasons. For example, NIST SP 800-57~\cite{nistsp80057} states that security levels below 80 bits are considered insecure for long-term use, with 112 bits as the current minimum and 128 bits recommended for cryptographic keys. Standards that govern random number generators (RNGs) are stricter still. NIST SP 800-90B~\cite{turan2018recommendation} effectively requires $\approx$1 bit of min-entropy per output bit ($\epsilon \leq 2^{-32}$), while AIS 20/31~\cite{bsi2024ais31} mandates over 240 and 250 bits of min-entropy and Shannon entropy respectively for an RNG's internal state. %As we will demonstrate, the min-entropy of even complex combinations of mobile sensors falls well below these thresholds.

In this paper, we analyse the entropy of 25 mobile sensors, which is substantially more than prior work---\cite{voris2011accelerometers} (1 sensor),~\cite{hennebert2013entropy} (10),~\cite{krhovjak2007sources} (2),~\cite{lv2020analysis} (3)---across multiple public datasets. Our results reveal widespread weaknesses, both corroborating and expanding on existing results. Commodity sensors exhibit pronounced biases and dependencies, yielding on average 3.408--4.483 bits of min-entropy and 5.584--9.266 bits of Shannon entropy per sample across datasets. Combining multiple sensors confers only modest gains: we explore how inter-sensor correlations and statistical bias contribute to a divergence between average- and worst-case entropy, with min-entropy typically $\approx$40–75\% below Shannon entropy for the same signals. Even when combining over 20 sensors simultaneously, the (joint) min-entropy remains below $\approx$22 bits, far below the thresholds considered secure by the aforementioned standards. In practical terms, schemes that treat these measurements as secrets expose a guess space small enough to be enumerable by an adversary. Our contributions are as follows:

\begin{itemize} 
\item We introduce the first systematic approach to evaluating sensor entropy across such a comprehensive range of modalities and datasets using various entropy metrics (max, Shannon, collision, and min-entropy).
\item We empirically demonstrate how inter-sensor correlations and biases erode entropy, casting doubt on the proposition that using multiple sensors adds substantially to security. 
\item We show how the collapse in worst-case entropy opens the door to attacks that exhaustively enumerate the space of possible measurements. This raises fundamental concerns with respect to security applications that rely on signals from single and multiple mobile sensors. Consequently, we do not recommend relying entirely on mobile sensors as entropy sources for security-critical applications.
\end{itemize}

The rest of this paper is organised as follows: \S\ref{sec:background} discusses sensor-based security mechanisms and established entropy metrics. \S\ref{sec:design} explains our experiment design for entropy estimation, including the threat model and dataset selection. \S\ref{sec:entropy-analysis} presents empirical results of our analyses and \S\ref{sec:security-evaluation} discusses the implications for system design. We conclude in \S\ref{sec:conc} with recommendations for further work. Our analysis work is released publicly to foster future research.\footnote{\url{https://github.com/cgshep/entropy-collapse-mobile-sensors}\label{fn:url}}

\section{Background}
\label{sec:background}

This section discusses sensor-based applications and existing approaches, and formalises the entropy metrics underpinning our subsequent analyses.

\subsection{Related Work}

Mobile sensing has been widely explored for security tasks in three broad families: (1) \emph{proximity detection} and \emph{relay-attack mitigation}, where co-presence is inferred from shared motion or ambient context~\cite{halevi2012secure,mehrnezhad2015tap,shepherd2017applicability,gurulian2017effectiveness,markantonakis2024using,gurulian2018good,shrestha2014drone}; (2) \emph{device pairing}, where two devices derive evidence of co-location to establish a secure association or shared secret~\cite{mayrhofer2009shake,pan2018universense,xu2021key,mayrhofer2021adversary}; and (3) \emph{continuous authentication}, which aims to verify users passively during normal interaction~\cite{riva2012progressive,shi2011senguard,miettinen2014conxsense,li2013unobservable,patel2016continuous,mekruksavanich2021deep,li2018sensor,micallef2015aren}. Representative designs include tap-to-pair schemes using accelerometers~\cite{mehrnezhad2015tap}, induced shared vibrations~\cite{gurulian2018good}, and ambient-signal approaches that leverage temperature, humidity, magnetic field, or other environmental cues~\cite{shrestha2014drone,markantonakis2024using}. Proposals commonly combine sensor modalities to gather additional evidence of co-presence; for example, using accelerometer and gyroscope measurements to capture shared device motion in addition to environmental readings~\cite{mayrhofer2009shake,pan2018universense,alzubaidi2016authentication,xu2021key}.

A common pipeline---documented across proximity detection, pairing, and continuous authentication---collects data from $N$ participants; derives time- or frequency-domain features (e.g.\ cross-correlation, spectral energy, and Hamming, Euclidean, and mean absolute distances); learns a decision rule (thresholds or supervised models such as SVMs, Random Forests, Na\"{i}ve Bayes); and reports FPR/FNR, EER, ROC/AUC, accuracy, or F1. This template appears in numerous systems, including~\cite{mehrnezhad2015tap,shrestha2014drone,markantonakis2024using,shepherd2017applicability,riva2012progressive,li2013unobservable,miettinen2014conxsense,patel2016continuous,mekruksavanich2021deep,gurulian2018good}. Across the above application areas, effectiveness is usually inferred from \emph{discriminative} performance, i.e.\ the system's performance in classifying illegitimate vs.\ legitimate samples.  (For a comprehensive treatment of the model evaluation metrics used by existing mobile sensor-based security systems, the reader is referred to the surveys by Abuhamad et al.~\cite{abuhamad2020sensor}, Alzubaidi and Kalita~\cite{alzubaidi2016authentication}, and Mayrhofer and Sigg~\cite{mayrhofer2021adversary}). Our key observation is that model-performance metrics are typically decoupled from the \emph{entropy} of the underlying data sources upon which such models are based.  Some proposals have attempted to assess the entropy of sensor signals within the context of a security mechanism, but this represents a small minority of the literature. Specifically, T2Pair by Li et~al.~\cite{li2020t2pair} reports 32.3--38.5 bits of Shannon entropy, and a refinement by Wu et~al.~\cite{wu2024t2pair} reports 51--54 bits. We are aware of no work that applies min-entropy as a worst-case measure of unpredictability~\cite{turan2018recommendation,bsi2024ais31,buller2016estimating,alzubaidi2016authentication,xu2021key,mayrhofer2021adversary}. Adjacent evidence also highlights practical limits under tight timing constraints. For NFC transactions, EMV specifies $\approx$500\,ms for proximity checks. In that setting, Markantonakis et~al.~\cite{markantonakis2024using} (building on~\cite{gurulian2017effectiveness,shepherd2017applicability}) report EERs of 0.179--0.246 across sensor combinations, concluding that commodity sensors are unsuitable for reliable proximity and relay detection without incurring usability or security issues. This points to wider fragility issues with sensor-based security schemes in practice. %Taken together, the literature demonstrates largely functional feasibility but leaves open the core question of how unpredictable the underlying measurements really are, especially under multimodal combinations. This motivates our entropy-centric evaluation in the remainder of the paper.
% We address this gap by analysing the entropy of a wide range of mobile sensors, both in isolation and in combination.

Some studies have attempted to measure the entropy of single mobile sensors. Voris et al.~\cite{voris2011accelerometers} evaluated accelerometers as true RNGs (TRNGs) on a WISP RFID tag and Nokia N97 phone. The authors find that min-entropy is proportional to the motion applied to the device, with stationary movement having the lowest min-entropy. Intrinsic noise from the sensor circuitry and seismic activity, and the sampling rate of its analog-to-digital converter (ADC), are found to influence entropy generation. Min-entropy values of 3.1--11.4 bits were measured depending on the movement of the accelerometer. Lv et al.~\cite{lv2020analysis} analysed three mobile sensors---a triaxial accelerometer, gyroscope, and magnetometer---on an undisclosed Xiaomi Redmi smartphone. Min-entropy values of 0.593--5.876 are reported, depending on the modality and the entropy estimation method.  Krhovj\'{a}k et al.~\cite{krhovjak2007sources} examined image and audio data collected from mobile phone cameras and microphones, respectively, on a Nokia N73 and E-Ten X500
and M700 phones. The work reports Shannon entropies of 2.9 (microphone) and 2.408--5.376 (camera), and min-entropies of 0.5 (microphone) and 0.754--3.928 (camera). Hennebert et al.~\cite{hennebert2013entropy} analysed 10 sensors on two wireless sensor monitors: a TI eZ430-RF2500 and a Zolertia Z1. Min-entropy values of 0--7.85 are given, with motion sensors, e.g.\ accelerometer and vibration sensors, producing the highest entropy.

Sensors have also been suggested as entropy sources within low-cost RNG designs for mobile devices. Suciu et al.~\cite{suciu2011unpredictable} proposed combining GPS along with accelerometer, gyroscope and orientation sensors on an HTC Google Nexus One, passing NIST SP 800-22 tests~\cite{rukhin2001nist}, but without precise entropy estimates.  Wallace et al.~\cite{wallace2016toward} explored a multi-sensor RNG design (accelerometer, gyroscope, microphone, WiFi, GPS, camera) with a non-standard entropy evaluation using 37 Android devices. Unlike dedicated TRNGs (e.g.\ ring oscillators, Johnson–Nyquist thermal noise, quantum sources~\cite{hurley2020quantum}), mobile sensors depend heavily on user behaviour and environmental factors, introducing biases and correlations absent in controlled entropy sources. Existing sensor-based mechanisms largely overlook these dynamics, relying principally on heuristic or model evaluation metrics that do not account well for skewed distributions and other biases affecting predictability.  In contrast, proposals of new entropy sources and authentication mechanisms commonly rely on worst-case metrics, such as min-entropy as stipulated in NIST SP 800-90B and AIS 20/31~\cite{mai2017guessability,uellenbeck2013quantifying,hurley2020quantum,turan2018recommendation,bsi2024ais31}. While several studies have investigated the entropy of specific sensors, a direct comparison is often difficult due to varying methodologies and contexts. Table~\ref{tab:prior_analyses} provides a structured summary of key prior works, contrasting the sensors analysed, entropy metrics used, and the reported entropy ranges.
 Our study bridges the gap by systematically evaluating entropy across modalities and datasets.

\begin{table}
\centering
\caption{Comparative summary of prior sensor entropy analyses.}
\resizebox{\linewidth}{!}{%
\begin{threeparttable}
\begin{tabular}{rllcp{6cm}}
\toprule
\textbf{Reference} & \textbf{Sensors}\tnote{a} & \textbf{Metrics}\tnote{b} & \textbf{Results (bits)} & \textbf{Context / Use-Case} \\ 
\midrule
Voris et al.~\cite{voris2011accelerometers} & Acc & $H_\infty$ & 3.1–11.4 & True random number generation \\
Hennebert et al.~\cite{hennebert2013entropy} & 10 (Various) & $H_\infty$ & 0--7.85 & Harvesting entropy from physical mobile sensors \\
Lv et al.~\cite{lv2020analysis} & Acc, Gyro, Mag & $H_\infty$ & 0.593--5.876 & Entropy source analysis on smartphone \\
Krhovj\'{a}k et al.~\cite{krhovjak2007sources} & Cam, Mic & $H_{\{1,\infty\}}$ & 0.5--5.376 & Randomness sources in mobile devices \\
Li et al.~\cite{li2020t2pair} & Mic, Light & $H_1$ & 32.3--38.5 & Zero-interaction device pairing \\
Wu et al.~\cite{wu2024t2pair} & Mic, Light & $H_1$ & 51--54 &Zero-interaction device pairing  \\ 
\midrule
\textbf{This work} & \textbf{25} (Various) & \textbf{$H_{\{0,1,2,\infty\}}$} & \textbf{8.1--23.9 ($H_\infty$)} & \textbf{Systematic entropy analysis, including sensor combinations, for mobile device sensors} \\\bottomrule
\end{tabular}
\begin{tablenotes}
    \item[a] Acc: Accelerometer; Gyro: Gyroscope; Mag: Magnetometer; Cam: Camera; Mic: Microphone.
    \item[b] $H_0$: Max; $H_1$: Shannon; $H_2$: Collision; $H_\infty$: Min-entropy.
\end{tablenotes}
\end{threeparttable}
}
\label{tab:prior_analyses}
\end{table}

\subsection{Definitions}
\label{sec:defs}
We use the following definitions and notation throughout this work. Furthermore, unless stated otherwise, all logarithms are base 2 and entropies are reported in bits.

\begin{definition}[R\'{e}nyi Entropy]
Let \(X\) be a discrete random variable taking values in a set \(\mathcal{X}\) with probability mass function \(p(x)\). 
The \emph{R\'{e}nyi entropy} of order \(\alpha\) (\(\alpha > 0\), \(\alpha \neq 1\)) is defined as
\begin{equation}
H_{\alpha}(X)
\,=\, \frac{1}{1 - \alpha} \,\log\!\Biggl(\sum_{x \in \mathcal{X}} p(x)^{\alpha}\Biggr).
\label{eq:renyi_entropy_general}
\end{equation}
\end{definition}

We use four cases of $\alpha$ that are widely used in the literature. First, the \emph{Hartley (Max) Entropy} (Eq.~\ref{eq:renyi_H0}) is the logarithm of the number of possible outcomes with non-zero probability and serves as an upper bound.
\begin{equation}
H_{0}(X) \equiv \lim_{\alpha \to 0}H_\alpha(X)
\,=\, \log \Bigl|\bigl\{\,x \in \mathcal{X}: p(x) > 0 \bigr\}\Bigr|.
\label{eq:renyi_H0}
\end{equation}

Second, \emph{Shannon entropy} (Eq.~\ref{eq:renyi_H1}) is the classical definition of entropy in information theory, corresponding to the limit of \(H_{\alpha}\) at \(\alpha \to 1\).
 
\begin{equation}
H_{1}(X) \equiv \lim_{\alpha \to 1}H_\alpha(X)
\,=\, - \sum_{x \in \mathcal{X}} p(x)\,\log p(x).
\label{eq:renyi_H1}
\end{equation}

Third, the \emph{collision entropy} (\(\alpha = 2\)) quantifies the probability of drawing the same value twice from \(X\):
\begin{equation}
H_{2}(X)
\,=\, -\log \Bigl(\sum_{x \in \mathcal{X}} p(x)^{2}\Bigr).
\label{eq:renyi_H2}
\end{equation}

We note that $H_1$ provides an average-case measure of uncertainty. It takes into account the entire distribution of outcomes; however, an adversary may only need to guess the most likely event to gain an advantage. To account for this, \emph{min-entropy} (Eq.~\ref{eq:renyi_Hinf}) is used as a conservative, worse-case metric, accounting for the least favorable distribution of outcomes; it is the value in the limit \(\alpha \to \infty\):
\begin{equation}
H_{\infty}(X) \equiv \lim_{\alpha \to \infty}H_\alpha(X)
\,=\, -\log \Bigl(\max_{x \in \mathcal{X}} p(x)\Bigr).
\label{eq:renyi_Hinf}
\end{equation}

$H_\alpha$ is a non-increasing function of $\alpha$, i.e.\ $H_{\infty}(X) \leq \dots \leq H_1(X)\leq H_0(X)$. $H_{\infty}(X)$ focuses on the single most likely outcome, providing a strictly tighter (and generally minimum) bound on uncertainty. We observe that min-entropy ensures that even the most skewed probability distributions still meet the required security guarantees; indeed, it is a recommended method for assessing entropy sources in modern standards~\cite{turan2018recommendation,bsi2024ais31}. We also rely on joint entropy for assessing the entropy of multiple random variables.

\begin{definition}[Joint R\'{e}nyi Entropy]
Let \(X_1, X_2, \dots, X_n\) be discrete random variables that jointly take values in
\(\mathcal{X}_1 \times \mathcal{X}_2 \times \cdots \times \mathcal{X}_n\),
with joint probability mass function
\(p(x_1, x_2, \dots, x_n)\).
The \emph{R\'{e}nyi entropy} of order \(\alpha\) (\(\alpha > 0, \alpha \neq 1\)) for these \(n\) variables is defined as the following, where the sum is taken over all \((x_1,\dots,x_n)\) in
\(\mathcal{X}_1 \times \mathcal{X}_2 \times \cdots \times \mathcal{X}_n\):
\begin{equation}
H_{\alpha}(X_1, X_2, \dots, X_n)
\,=\, \frac{1}{1 - \alpha}\,\log\Biggl(
  \sum_{(x_1,x_2,\dots,x_n)} p(x_1, x_2, \dots, x_n)^{\alpha}
\Biggr)
\label{eq:multivariate_renyi_entropy}
\end{equation}

\end{definition}

In the limit \(\alpha \to 1\), then \(H_{\alpha}(X_1, X_2, \dots, X_n)\) converges to the classical joint Shannon entropy of these \(n\) variables. To support the later discussion on Chow-Liu trees---our approach to joint entropy estimation---we define the Kullback-Leibler divergence.

\begin{definition}[Kullback-Leibler (KL) Divergence]\label{def:kl} Given a true probability distribution \( P(x) \) of a random variable and an approximate or reference distribution \( Q(x) \), the KL divergence is defined as follows:
\begin{equation}
    D_{KL}(P\;||\; Q) = \sum_{x \in \mathcal{X}}P(x)\log\Big(\frac{P(x)}{Q(x)}\Big)
\end{equation}
\end{definition}

\section{Experiment Design}
\label{sec:design}

\subsection{Threat Model and Assumptions}

We consider an adversary aiming to compromise sensor-based security schemes (e.g.\ key generation, proximity detection, and continuous authentication). The adversary may gather extensive statistical data about how smartphone sensors behave in everyday usage; for instance, from widely available open datasets. We assume the attacker focuses on predicting or guessing the sensor outputs by prioritising the most probable values first, exploiting any biases in the distribution of sensor measurements. As such, they may resort to exhaustive enumeration of the measurement space if the target source's entropy is low enough.
We exclude capabilities such as fault injection and other hardware attacks (see~\cite{shepherd2021physical}) as well as side-channel attacks such as cross-device correlation attacks~\cite{dautov2019effects}. While those could further reduce the effective entropy space---say, by inducing errors in the output values of sensing hardware---we regard them as outside the scope of this work. This threat model represents an adversary who can capitalise on statistical biases in sensor data without directly comprising the device physically. Our goal is to evaluate whether sensor data distributions offer sufficient entropy to resist attacks that search the space of sensor measurement values informed by their statistical properties.

\subsection{High-level Methodology}

 In this study, we focus on estimating population-level (``global'') sensor distributions by pooling all users within a dataset. This choice reflects typical deployments in which systems must serve previously unseen users (e.g.\ first-time proximity checks, payment defence mechanisms, or ad-hoc device pairing) and thus cannot assume per-user enrollment or stable user-specific priors.  Our approach involves five main stages:

\begin{enumerate}
    \item \emph{Data acquisition.} We collect large-scale sensor readings from public datasets spanning diverse activities and device usage. Modalities include motion, environmental, and orientation sensors (details are given at the end of this section).
    \item \emph{Global distributions.}  We merge sensor readings into a single, global distribution for each sensor modality in each dataset. For sensors that are inherently discrete or quantised (e.g., integer output ranges), we simply count occurrences. For sensors that produce (quasi-)continuous values, we rely on quantisation using Freedman–Diaconis binning to partition the output space and approximate an empirical probability mass function.

    \item \emph{Single-sensor entropy analysis.} From these global distributions, we compute max, Shannon and collusion entropies to measure the best- and average-case uncertainties of sensor outputs, along with the min-entropy to characterise the worst-case unpredictability.
    \item \emph{Multimodal entropy estimation.} Many real-world proposals combine multiple sensor streams to purportedly increase security. To assess the impact on worst-case unpredictability, we use Chow-Liu trees to approximate the joint distributions of different sensor modalities. This allows us to estimate higher-dimensional entropies without incurring prohibitive computational costs (discused in \S\ref{sec:multimodal}).
    \item \emph{Evaluation.} Finally, we interpret the resulting entropy measures, focusing on whether sensor outputs remain sufficiently unpredictable against an informed adversary. We compare single-sensor versus multi-sensor scenarios to verify if combining modalities truly alleviates biases or simply adds redundant data susceptible to predictability concerns.

\end{enumerate}

Throughout, we note the constraints of NIST SP~800-90B, SP~800-22~\cite{turan2018recommendation}, and AIS~20/31~\cite{bsi2024ais31} when applied to multi-sensor streams~\cite{buller2016estimating,lv2020analysis}. These frameworks target univariate sources rather than joint entropy of complex multivariate processes, which is the focus of our work. (For example, NIST SP 800-90B focusses on assessing entropy sources with reduced, i.e.\ 8-bit, outputs~\cite{buller2016estimating,turan2018recommendation}). To scope suitable data, we searched IEEE DataPort, Google Scholar, Google Dataset Search, and GitHub. Several ostensibly “open” datasets were unavailable for download or had restrictive licensing~\cite{Mahbub_Btas2016_UMDAA02,stragapede2023behavepassdb,acien2021becaptcha}. We selected four datasets offering diverse contexts, consistent sampling, and documented modalities:
\begin{itemize}
    \item \emph{UCI-HAR}~\cite{anguita2013public}:  A widely referenced dataset for human activity recognition, comprising smartphone sensor recordings from multiple subjects performing daily activities. Data includes triaxial accelerometer and gyroscope signals.    
    \item \emph{University of Sussex--Huawei Locomotion (SHL)}~\cite{gjoreski2018university,wang2019enabling}: Sampled at 100 Hz from an Huawei Mate 9 smartphone. We use the publicly available SHL Preview dataset containing three recording days per user (59 hours total). For scope alignment with prior work, the handheld-phone subset is used.
    \item \emph{Relay}~\cite{markantonakis2024using}: Approximately 1{,}500 NFC-based contactless transactions recorded at 100\,Hz across several physical locations (e.g.\ caf\'{e}s), including accelerometer, gyroscope, and environmental readings taken from emulated payment scenarios.
    \item \emph{PerilZIS}~\cite{fomichev2019perils}: Measurements collected at 10\,Hz from a TI SensorTag, Samsung Galaxy~S6, and Samsung Galaxy Gear, spanning multiple zero-interaction security use cases in an office environment.
\end{itemize}

These datasets provide varied sensor types, user activities, and sampling rates, enabling us to observe how  biases and inter-sensor correlations manifest across scenarios. We summarise the datasets in Table~\ref{tab:datasets}. %Next, we describe the preprocessing and aggregation steps in forming the global distributions used in our analyses.

\begin{table}[t]
\centering
\footnotesize
\begin{threeparttable}
\caption{Summary of datasets.}
\label{tab:datasets}
\begin{tabular}{l|p{3.2cm}|p{3.4cm}|l|p{2.2cm}}
\toprule
\textbf{Dataset} & \textbf{Devices} & \textbf{Sensors}\tnote{a} & \textbf{Freq.} & \textbf{Context}\tnote{b} \\
\midrule
UCI-HAR~\cite{anguita2013public} & Samsung Galaxy S2 & Acc, Gyro & 50Hz & ADL; 30 users \\[0.25cm]
SHL~\cite{gjoreski2018university,wang2019enabling} & Huawei Mate 9 & Acc, Gyro, Mag, Grav, LinAcc, RotVec, Light, Temp, Pres, Alt & 100Hz & HAR; transport; 3 users; 59 hrs \\[0.8cm]
Relay~\cite{markantonakis2024using} & Google Nexus 5, 9; Samsung Galaxy S4, SGS5 & Acc, Gyro, Mag, RotVec, Grav, LinAcc, Light & 100Hz & NFC transactions; 252 users \\[0.5cm]
PerilZIS~\cite{fomichev2019perils} & TI SensorTag; Samsung Galaxy S6, Gear & Acc, Gyro, Mag, Light, Temp, Pres & 10Hz & ZIS; 4 users; 8 devices;  4250~hrs \\
\bottomrule
\end{tabular}

\begin{tablenotes}[flushleft]\footnotesize
\item[a] Acc: Accelerometer; Gyro: Gyroscope; Mag: Magnetometer; Grav: Gravity; LinAcc: Linear acceleration; RotVec: Rotation vector; Temp: Temperature; Pres: Pressure; Alt: Altitude.
\item[b] ADL: Activities of daily living; HAR: Human activity recognition; NFC: Near-field communication; ZIS: Zero-interaction security.
\end{tablenotes}
\end{threeparttable}
\end{table}
\section{Entropy Analysis}
\label{sec:entropy-analysis}

In this section, we analyse the entropy of sensor data on a single- and multi-modal basis. We begin by discussing the challenges in quantising sensor values for discrete-entropy calculations, then present our findings for our single- and multi-sensor analyses.

\subsection{Pre-processing}
\label{sec:preproc}
\begin{figure*}
    \centering
    \begin{subfigure}[t]{0.333\textwidth}
        \centering
        \includegraphics[width=\textwidth]{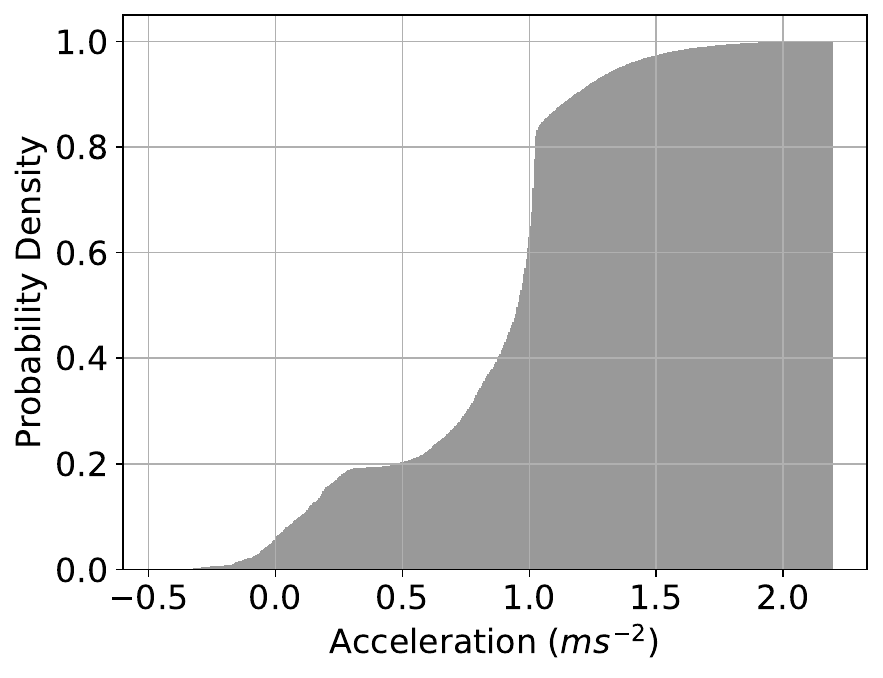}
        \caption{Acc.\ $x$ axis.}
        \label{subfig:acx}
    \end{subfigure}%
    \begin{subfigure}[t]{0.333\textwidth}
        \centering
        \includegraphics[width=\textwidth]{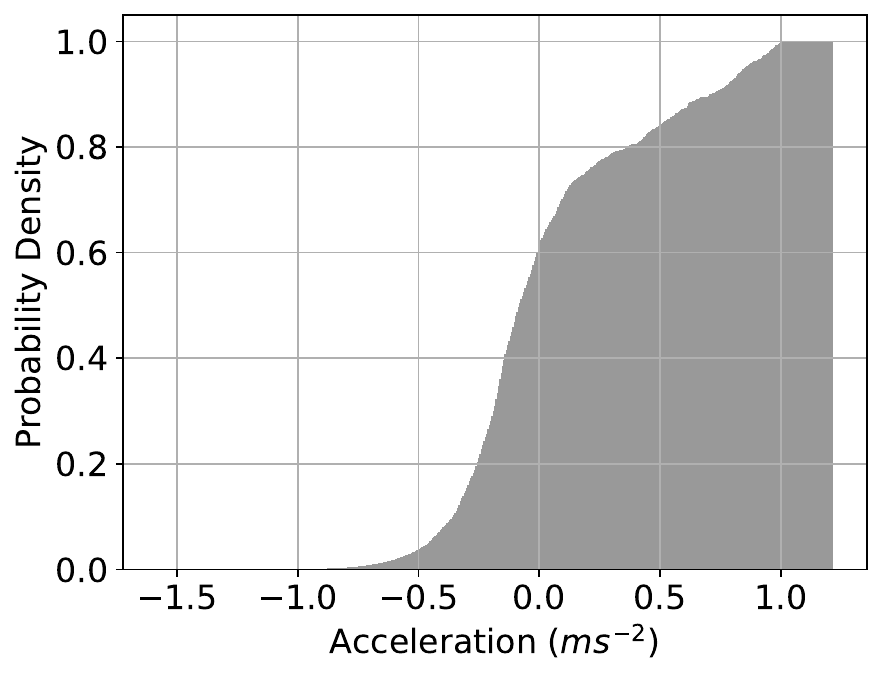}
        \caption{Acc.\ $y$ axis.}
        \label{subfig:acy}
    \end{subfigure}%
    \begin{subfigure}[t]{0.333\textwidth}
        \centering
        \includegraphics[width=\textwidth]{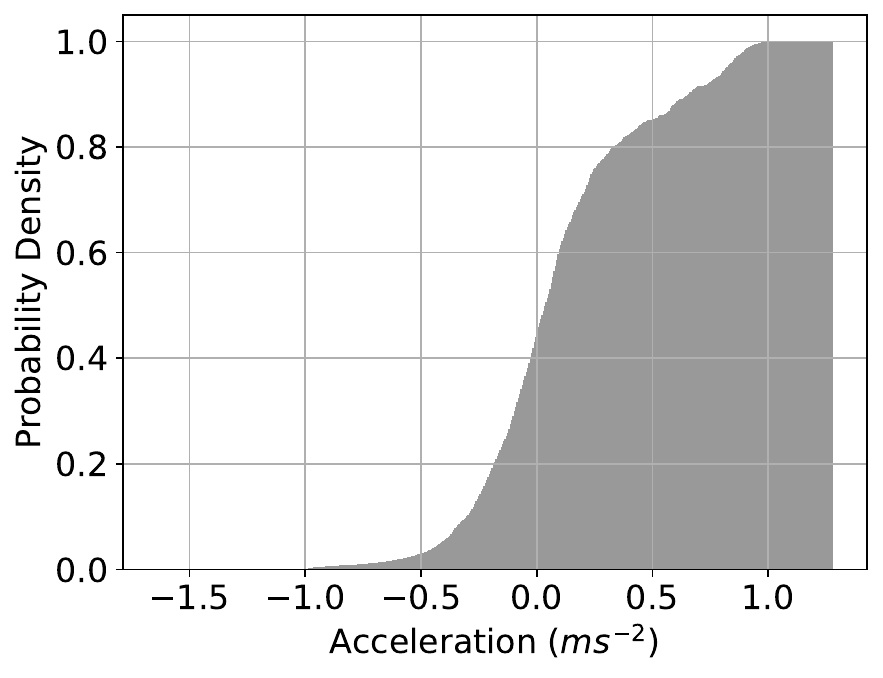}
        \caption{Acc.\ $z$ axis.}
        \label{subfig:acz}
    \end{subfigure}

    \begin{subfigure}[t]{0.333\textwidth}
        \centering
        \includegraphics[width=\textwidth]{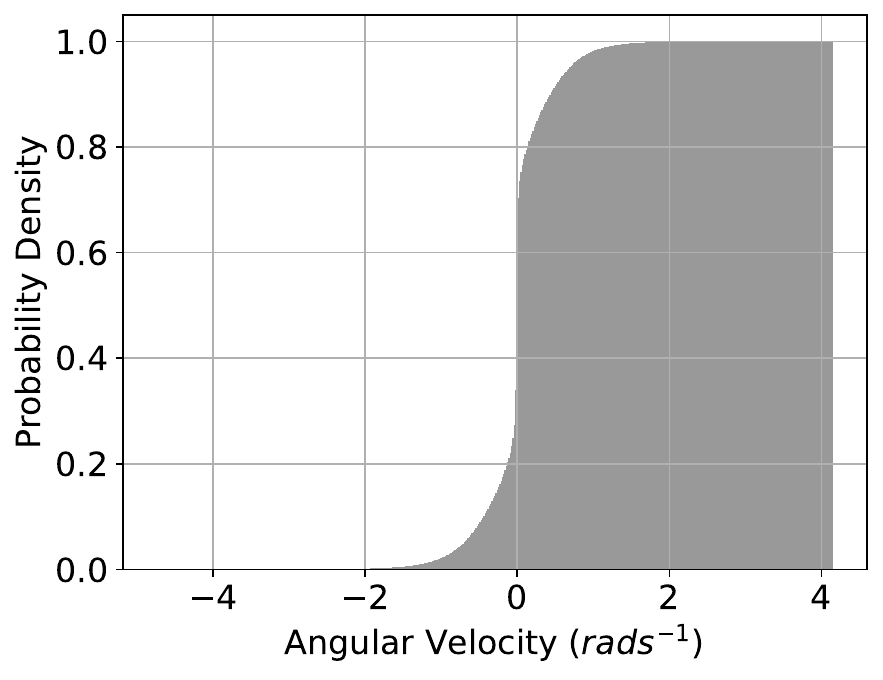}
        \caption{Gyro.\ $x$ axis.}
        \label{subfig:gyx}
    \end{subfigure}%
    \begin{subfigure}[t]{0.333\textwidth}
        \centering
        \includegraphics[width=\textwidth]{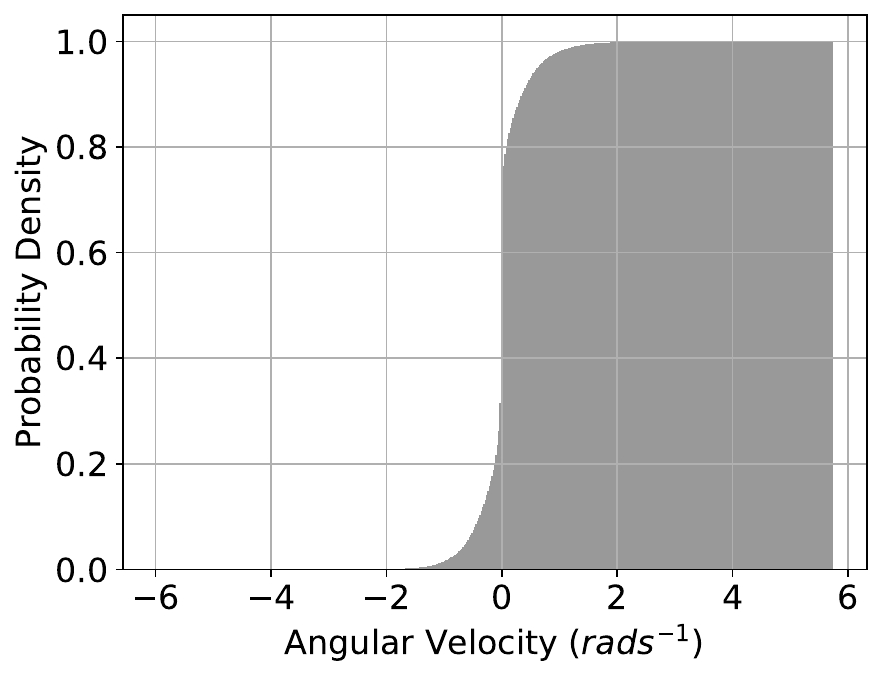}
        \caption{Gyro.\ $y$ axis.}
        \label{subfig:gyy}
    \end{subfigure}%
    \begin{subfigure}[t]{0.333\textwidth}
        \centering
        \includegraphics[width=\textwidth]{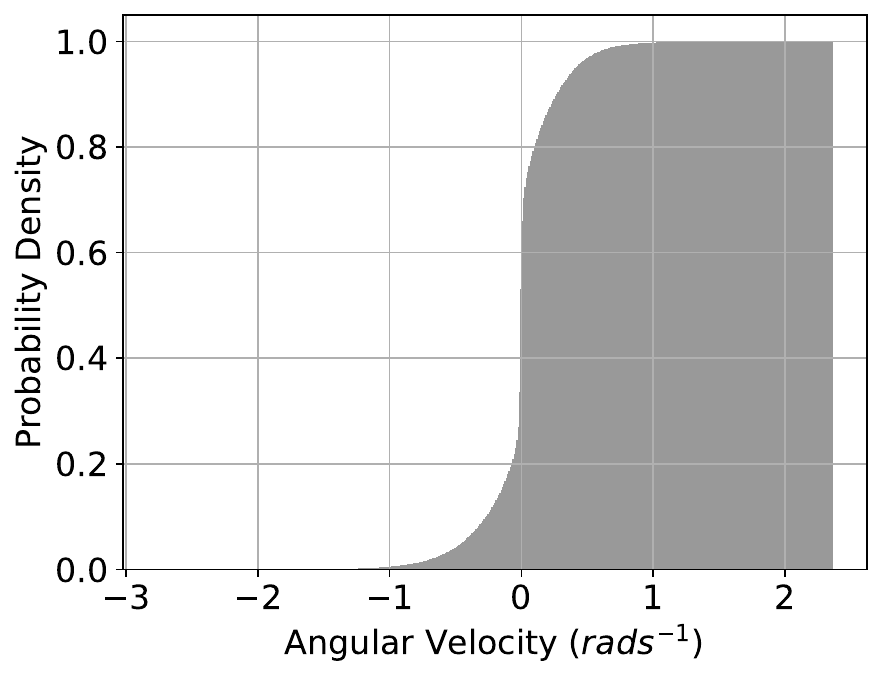}
        \caption{Gyro.\ $z$ axis.}
        \label{subfig:gyz}
    \end{subfigure}

    \begin{subfigure}[t]{0.333\textwidth}
        \centering
        \includegraphics[width=\textwidth]{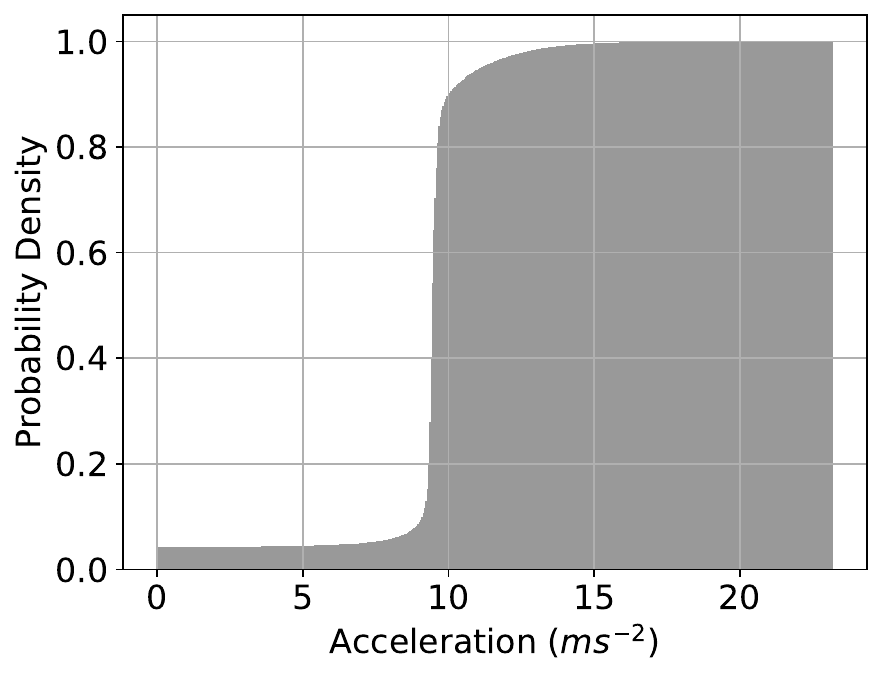}
        \caption{Accelerometer.}
        \label{subfig:acc}
    \end{subfigure}%
    \begin{subfigure}[t]{0.333\textwidth}
        \centering
        \includegraphics[width=\textwidth]{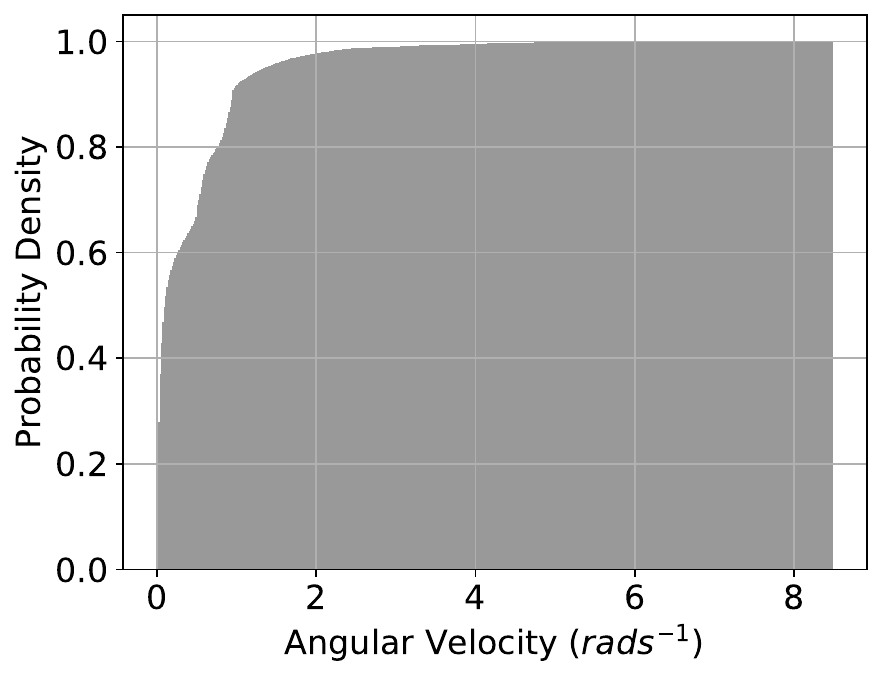}
        \caption{Gyroscope.}
        \label{subfig:gyro}
    \end{subfigure}%
    \begin{subfigure}[t]{0.333\textwidth}
        \centering
        \includegraphics[width=\textwidth]{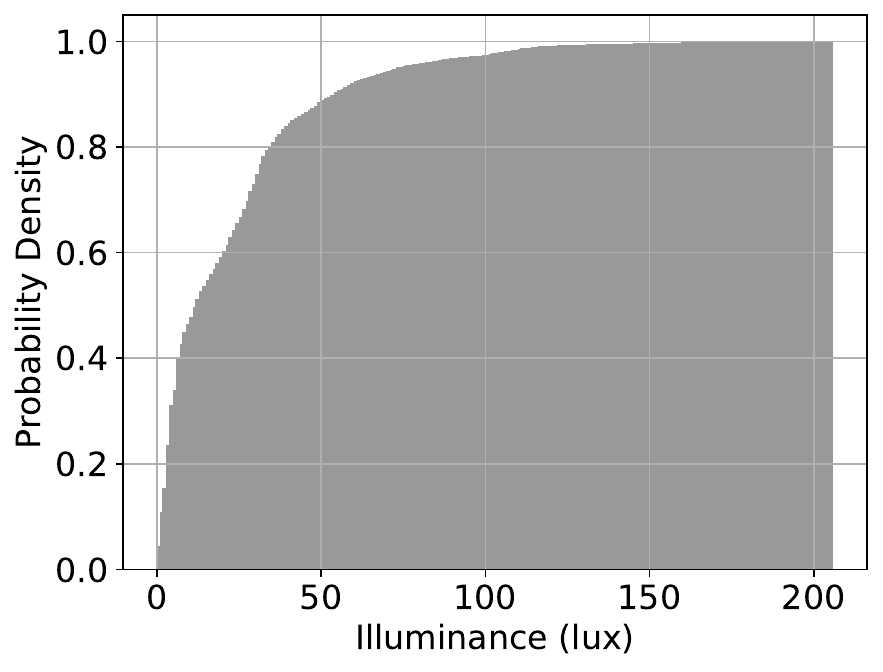}
        \caption{Light.}
        \label{subfig:light}
    \end{subfigure}

    \begin{subfigure}[t]{0.333\textwidth}
        \centering
        \includegraphics[width=\textwidth]{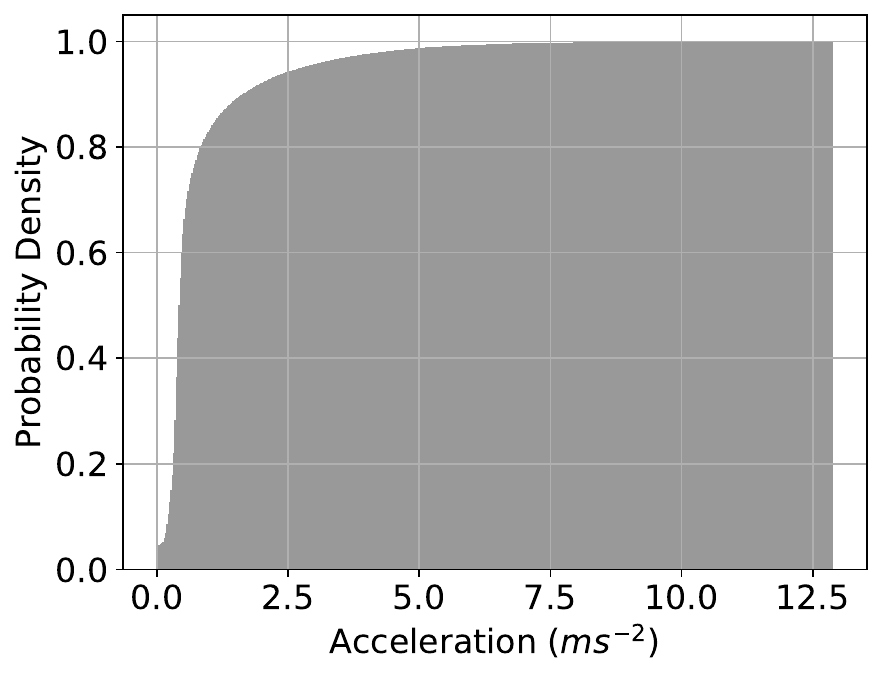}
        \caption{Linear Acceleration.}
        \label{subfig:linacc}
    \end{subfigure}%
    \begin{subfigure}[t]{0.333\textwidth}
        \centering
        \includegraphics[width=\textwidth]{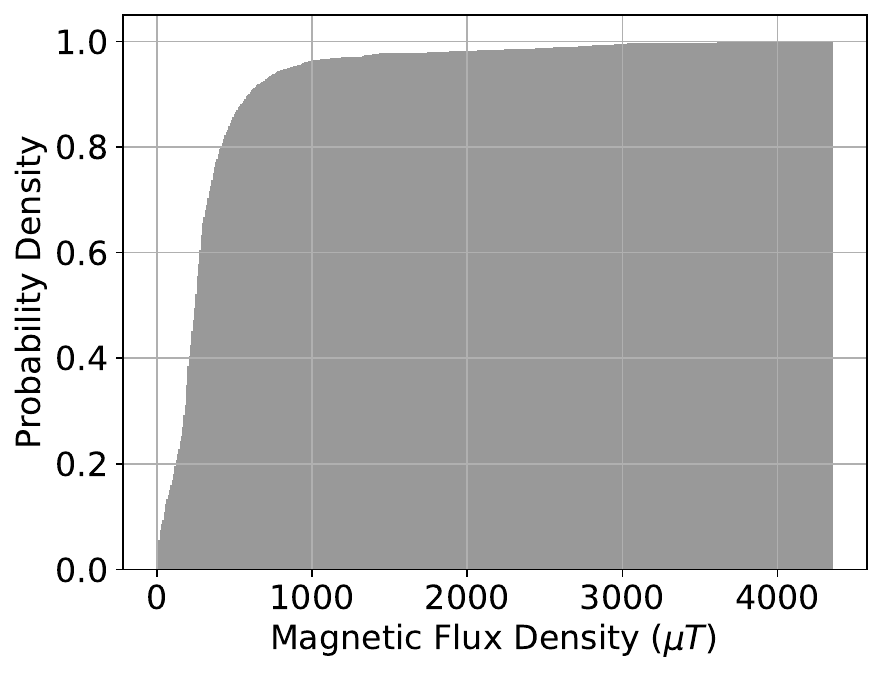}
        \caption{Magnetometer.}
        \label{subfig:mag}
    \end{subfigure}%
    \begin{subfigure}[t]{0.333\textwidth}
        \centering
        \includegraphics[width=\textwidth]{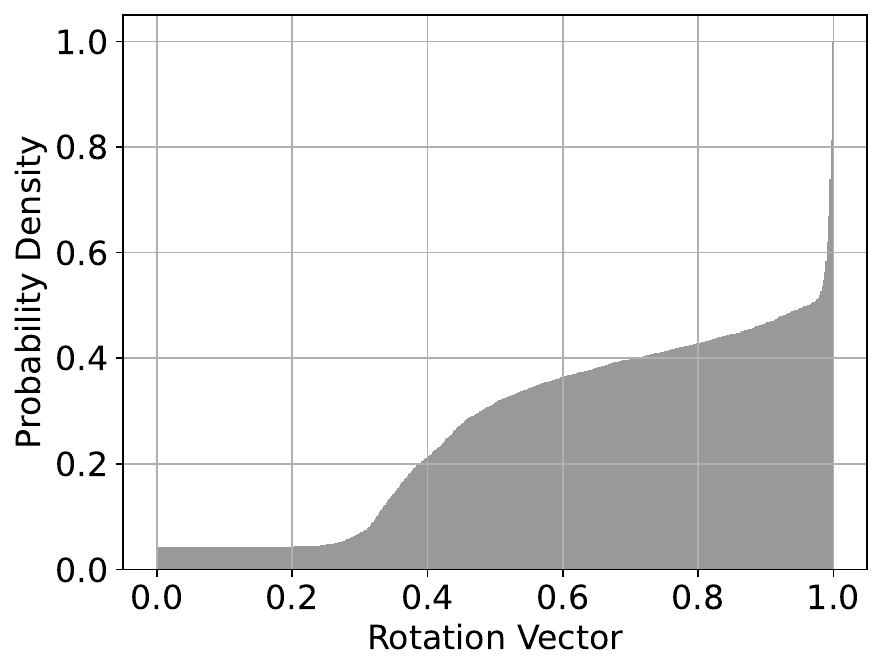}
        \caption{Rotation Vector.}
        \label{subfig:rv}
    \end{subfigure}
    \caption{Global sensor data CDFs -- UCI-HAR (a--f) and Relay (g--l) datasets.}
    \label{fig:acc-cdfs}
\end{figure*}

A crucial, yet underexplored, issue in prior work (e.g.~\cite{voris2011accelerometers,lv2020analysis,hennebert2013entropy,mai2017guessability}) is how to appropriately estimate entropy from quantised sensor outputs. Physical quantities such as linear acceleration or angular velocity are continuous in nature, even though modern sensors employ internal analog-to-digital conversion with finite resolutions. Notably, a sensor's advertised resolution---for instance, 12 bits in the widely used Bosch BMA400 mobile accelerometer~\cite{bosch_bma400}---does \emph{not} imply uniform histograms across its range of values. Everyday usage produces highly non-uniform biases and clustering, resulting in some measurements occurring more frequently than others. For example, \emph{UCI-HAR} data shows accelerometer readings concentrated in certain areas, with approximately 60\% of gyroscope readings concentrated around zero. As an illustration, Figure~\ref{fig:acc-cdfs} shows the cumulative probability distributions of some selected modalities across our datasets. We note that such skew and bias radically diminishes entropy compared to uniformly distributed values.

Another practical issue is the sparsity of sensor measurements at high resolutions. Treating every minute fluctuation e.g.\ $9.001\,\mathrm{m\,s^{-2}}$ vs.\ $9.002\,\mathrm{m\,s^{-2}}$) as distinct outcomes can inflate entropy estimates.  In real-world applications, it is the `similarity' between measurement signals that is considered useful in existing work; humans cannot reliably reproduce high-precision movements capable of utilising a sensor's digital resolution at, say, a $0.01\mathrm{ms^{-2}}$ scale. We therefore aggregate adjacent codes/values into bins to obtain reliable probability mass functions (pmfs). The choice of bin width is non-trivial: classical rules (e.g.\ Sturges, Scott, Doane) encode different assumptions about the underlying distribution, vary in computational cost, and differ in robustness to outliers. We adopt the Freedman--Diaconis (FD) rule, a distribution-agnostic, outlier-robust choice that adapts to sample size and spread,\footnote{Alternatively, a perceptual/actuation-driven scheme aligned with human reproducibility could be used; we defer this to future work.} with width $h$ given in Eq.~\ref{sec:freedman}, where $\mathrm{IQR}(x)$ is the interquartile range and $n$ the number of samples. Later, in \S\ref{sec:bin-sensitivity}, we show that FD falls within the plateau region as part of a bin sensitivity analysis, yielding stable estimates across datasets while avoiding oversmoothing and sparsity.

\begin{equation}
    h = 2 \cdot \frac{IQR(x)}{n^{1/3}}
    \label{sec:freedman}
\end{equation}

\subsection{Single Sensors}

\begin{table}
\renewcommand{\arraystretch}{2}
\centering
\caption{Single-sensor entropy values (in bits) for each dataset. Grey cells denote unavailable data for that dataset and modality.}
\resizebox{\linewidth}{!}{%
\label{tab:single-sensor-results}
\small
\begin{tabular}{@{}r|ccc|c|ccc|c|ccc|c|ccc|c@{}}
\toprule
 & \multicolumn{16}{c}{\textbf{Dataset}} \\
 & \multicolumn{4}{c|}{\textbf{UCI-HAR}} 
 & \multicolumn{4}{c|}{\textbf{SHL}}  
 & \multicolumn{4}{c|}{\textbf{Relay}*}  
 & \multicolumn{4}{c}{\textbf{PerilZIS}} \\
\midrule
\textbf{Sensor} 
 & \(H_{0}\) & \(H_{1}\) & \(H_{2}\) & \(H_{\infty}\) 
 & \(H_{0}\) & \(H_{1}\) & \(H_{2}\) & \(H_{\infty}\) 
 & \(H_{0}\) & \(H_{1}\) & \(H_{2}\) & \(H_{\infty}\) 
 & \(H_{0}\) & \(H_{1}\) & \(H_{2}\) & \(H_{\infty}\) \\
\midrule
Acc.X       
 & 8.488 & 7.080 & 5.876 & 3.729 
 & 11.557 & 8.732  & 7.487  & 4.543 
 & \missingcell & \missingcell & \missingcell & \missingcell 
 & 13.012 & 9.292  & 6.359  & 3.626 \\
Acc.Y       
 & 8.243 & 7.231 & 6.847 & 5.694 
 & 11.425 & 8.928  & 7.717  & 4.500 
 & \missingcell & \missingcell & \missingcell & \missingcell 
 & 9.549  & 5.873  & 4.483  & 2.889 \\
Acc.Z       
 & 8.455 & 7.397 & 7.069 & 6.020 
 & 10.428 & 7.627  & 6.366  & 3.785 
 & \missingcell & \missingcell & \missingcell & \missingcell 
 & 9.817  & 6.671  & 5.403  & 4.002 \\
Acc.Mag    
 & 8.895 & 6.284 & 4.819 & 3.489 
 & 14.583 & 10.136 & 8.710  & 6.435 
 & 10.145 & 6.843 & 5.808 & 4.538
 & 13.328 & 8.273  & 7.115  & 4.526 \\
\midrule
Gyro.X     
 & 8.683 & 5.430 & 3.504 & 1.929 
 & 15.024 & 10.532 & 8.107  & 4.993 
 & \missingcell & \missingcell & \missingcell & \missingcell 
 & 14.528 & 7.231  & 4.454  & 2.805 \\
Gyro.Y     
 & 8.439 & 5.023 & 3.461 & 2.300 
 & 15.085 & 10.283 & 7.601  & 4.827 
 & \missingcell & \missingcell & \missingcell & \missingcell 
 & 14.078 & 6.715  & 4.039  & 2.529 \\
Gyro.Z     
 & 8.714 & 5.675 & 3.948 & 2.363 
 & 15.281 & 10.070 & 5.708  & 3.083 
 & \missingcell & \missingcell & \missingcell & \missingcell 
 & 13.961 & 6.463  & 3.836  & 2.434 \\
Gyro.Mag  
 & 8.414 & 5.759 & 4.130 & 2.537 
 & 12.123 & 7.816  & 5.728  & 3.699 
 & 7.954 & 4.751 & 3.442 & 2.083 
 & 14.166 & 5.565  & 1.932  & 0.969 \\
\midrule
Mag.X   
 & \missingcell & \missingcell & \missingcell & \missingcell 
 & 12.845 & 8.840  & 8.386  & 6.374 
 & \missingcell & \missingcell & \missingcell & \missingcell 
 & 10.767 & 7.639  & 6.816  & 4.883 \\
Mag.Y   
 & \missingcell & \missingcell & \missingcell & \missingcell 
 & 12.263 & 8.737  & 8.314  & 6.223 
 & \missingcell & \missingcell & \missingcell & \missingcell 
 & 10.179 & 7.622  & 6.605  & 4.405 \\
Mag.Z   
 & \missingcell & \missingcell & \missingcell & \missingcell 
 & 12.516 & 8.586  & 8.217  & 6.228 
 & \missingcell & \missingcell & \missingcell & \missingcell 
 & 10.129 & 7.507  & 6.726  & 4.448 \\
Mag.Mag 
 & \missingcell & \missingcell & \missingcell & \missingcell 
 & 13.558 & 9.436  & 8.771  & 7.148 
 & 7.972 & 6.147 & 5.617 & 4.254
 & 10.293 & 7.329  & 6.489  & 4.454 \\
\midrule
Rot.\ Vec.\ 
 & \missingcell & \missingcell & \missingcell & \missingcell 
 & 8.725  & 7.721  & 5.970  & 3.220 
 & 5.000 & 3.307 & 1.965 & 1.021
 & \missingcell & \missingcell & \missingcell & \missingcell \\
\midrule
Grav.X     
 & \missingcell & \missingcell & \missingcell & \missingcell 
 & 9.014  & 8.482  & 7.266  & 4.299 
 & \missingcell & \missingcell & \missingcell & \missingcell 
 & \missingcell & \missingcell & \missingcell & \missingcell \\
Grav.Y     
 & \missingcell & \missingcell & \missingcell & \missingcell 
 & 9.338  & 8.770  & 7.602  & 4.453 
 & \missingcell & \missingcell & \missingcell & \missingcell 
 & \missingcell & \missingcell & \missingcell & \missingcell \\
Grav.Z     
 & \missingcell & \missingcell & \missingcell & \missingcell 
 & 8.180  & 7.193  & 5.418  & 3.036 
 & \missingcell & \missingcell & \missingcell & \missingcell 
 & \missingcell & \missingcell & \missingcell & \missingcell \\
Grav.Mag     
 & \missingcell & \missingcell & \missingcell & \missingcell 
 & 14.373 & 7.988  & 7.227  & 6.242 
 & 7.794 & 6.325 & 5.864 & 4.532
 & \missingcell & \missingcell & \missingcell & \missingcell \\
\midrule
LinAcc.X  
 & \missingcell & \missingcell & \missingcell & \missingcell 
 & 15.260 & 10.077 & 7.621  & 5.116 
 & \missingcell & \missingcell & \missingcell & \missingcell 
 & \missingcell & \missingcell & \missingcell & \missingcell \\
LinAcc.Y  
 & \missingcell & \missingcell & \missingcell & \missingcell 
 & 14.859 & 10.116 & 7.639  & 5.224 
 & \missingcell & \missingcell & \missingcell & \missingcell 
 & \missingcell & \missingcell & \missingcell & \missingcell \\
LinAcc.Z  
 & \missingcell & \missingcell & \missingcell & \missingcell 
 & 14.377 & 9.951  & 7.605  & 4.543 
 & \missingcell & \missingcell & \missingcell & \missingcell 
 & \missingcell & \missingcell & \missingcell & \missingcell \\
LinAcc.Mag  
 & \missingcell & \missingcell & \missingcell & \missingcell 
 & 12.777 & 7.968  & 5.752  & 3.420 
 & 9.175 & 6.385 & 5.424 & 4.222
 & \missingcell & \missingcell & \missingcell & \missingcell \\
\midrule
Light       
 & \missingcell & \missingcell & \missingcell & \missingcell 
 & \missingcell & \missingcell & \missingcell & \missingcell 
 & 7.200& 5.331 & 4.507& 3.206
 & 12.152 & 7.940  & 7.137  & 4.552 \\
Humidity    
 & \missingcell & \missingcell & \missingcell & \missingcell 
 & \missingcell & \missingcell & \missingcell & \missingcell 
 & \missingcell & \missingcell & \missingcell & \missingcell 
 & 7.943  & 7.048  & 6.774  & 5.546 \\
Temp.       
 & \missingcell & \missingcell & \missingcell & \missingcell 
 & 7.295  & 4.753  & 2.611  & 1.332 
 & \missingcell & \missingcell & \missingcell & \missingcell 
 & 8.484  & 7.416  & 6.941  & 5.449 \\
Pressure    
 & \missingcell & \missingcell & \missingcell & \missingcell 
 & 9.461  & 8.170  & 7.723  & 6.237 
 & \missingcell & \missingcell & \missingcell & \missingcell 
 & 8.044  & 7.006  & 6.370  & 5.073 \\
\midrule 
\textbf{Mean} 
 & 8.541 & 6.235 & 4.957 & 3.508 
 & 13.188 & 9.266 & 7.178 & 4.483 
 & 7.891 & 5.584 & 4.661 & 3.408 
 & 11.277 & 7.224 & 5.717 & 3.912 \\
\textbf{S.D.} 
 & 0.207 & 0.904 & 1.461 & 1.574 
 & 1.993 & 1.148 & 1.115 & 1.018 
 & 1.612 & 1.227 & 1.474 & 1.379 
 & 2.312 & 0.900 & 1.526 & 1.266 \\\bottomrule
\end{tabular}
}
\end{table}

Given the biases discussed above, it is inevitable that some sensor readings will be highly predictable. To quantify this, we calculate individual-sensor entropies across datasets. The results are given in Table~\ref{tab:single-sensor-results}. For multi-dimensional modalities (e.g.\ triaxial accelerometer or gyroscope), these are split into separate axes following Voris et al.~\cite{voris2011accelerometers}. This approach allows for a fine-grained analysis of directional biases, as movement along one axis may be more constrained or predictable than others in typical usage scenarios.  We note that, in the Relay dataset, the data for individual $x$, $y$ and $z$ components are not given for the accelerometer, gyroscope, and magnetometer sensors. Rather, the authors have already preprocessed triaxial data into its vector magnitudes, i.e.\ $\textbf{v} = \sqrt{x^2 + y^2 + z^2}$.  This magnitude represents the overall intensity of motion, irrespective of direction. We give this as ``X.Mag'' for a given sensor X. For the sake of completeness and consistency, we compute the magnitude for other datasets, where applicable, and report the entropy values for this synthetic modality.

Several clear patterns emerge from Table~\ref{tab:single-sensor-results}. Some sensors, such as certain accelerometer axes in \emph{SHL} or \emph{PerilZIS}, exhibit moderate min-entropies of 4--6 bits. Other sensors, particularly gyroscope axes (see \emph{UCI-HAR}) show values below 3 bits, indicating high predictability. Shannon entropy values (\(H_1\)) can be fairly high (up to 10 bits in some cases), whereas min-entropy (\(H_{\infty}\)) is often much lower. This gap reflects distributions where a few outcomes dominate, driving worst-case unpredictability down even if the average-case picture is more favorable. In general, the results confirm that data from individual sensors do not provide sufficient min-entropy for robust security on their own. In the next section, we examine whether combining multiple modalities can meaningfully increase this worst-case unpredictability or whether correlated biases persist across different sensor streams.

\subsection{Multi-modal Sensors}
\label{sec:multimodal}

Several sensor-based proposals~\cite{mehrnezhad2015tap,truong2014comparing,markantonakis2024using,shrestha2014drone,shrestha2018sensor} assert that combining multiple sensor modalities bolsters security, on the intuition that an adversary must accurately predict several data streams rather than a single one. An initial, na\"{i}ve approach may attempt to model this by adding the Shannon entropies from individual sensors, benefitting from the relation $H(X_1, \ldots, X_n) = \sum_{i=1}^n H(X_i)$. However, this requires that $X_i$ are \emph{statistically independent.} In reality, mobile sensors often exhibit strong dependencies; for instance, the rotation vector, gravity, and linear acceleration sensors are frequently derived in software from the accelerometer and gyroscope on consumer devices~\cite{android_motion_sensors}. As a result, these modalities \emph{cannot} be treated as independent variables.

\begin{figure*}[t!]
    \centering
    \begin{subfigure}[t]{0.45\textwidth}
        \centering
        \includegraphics[width=\textwidth]{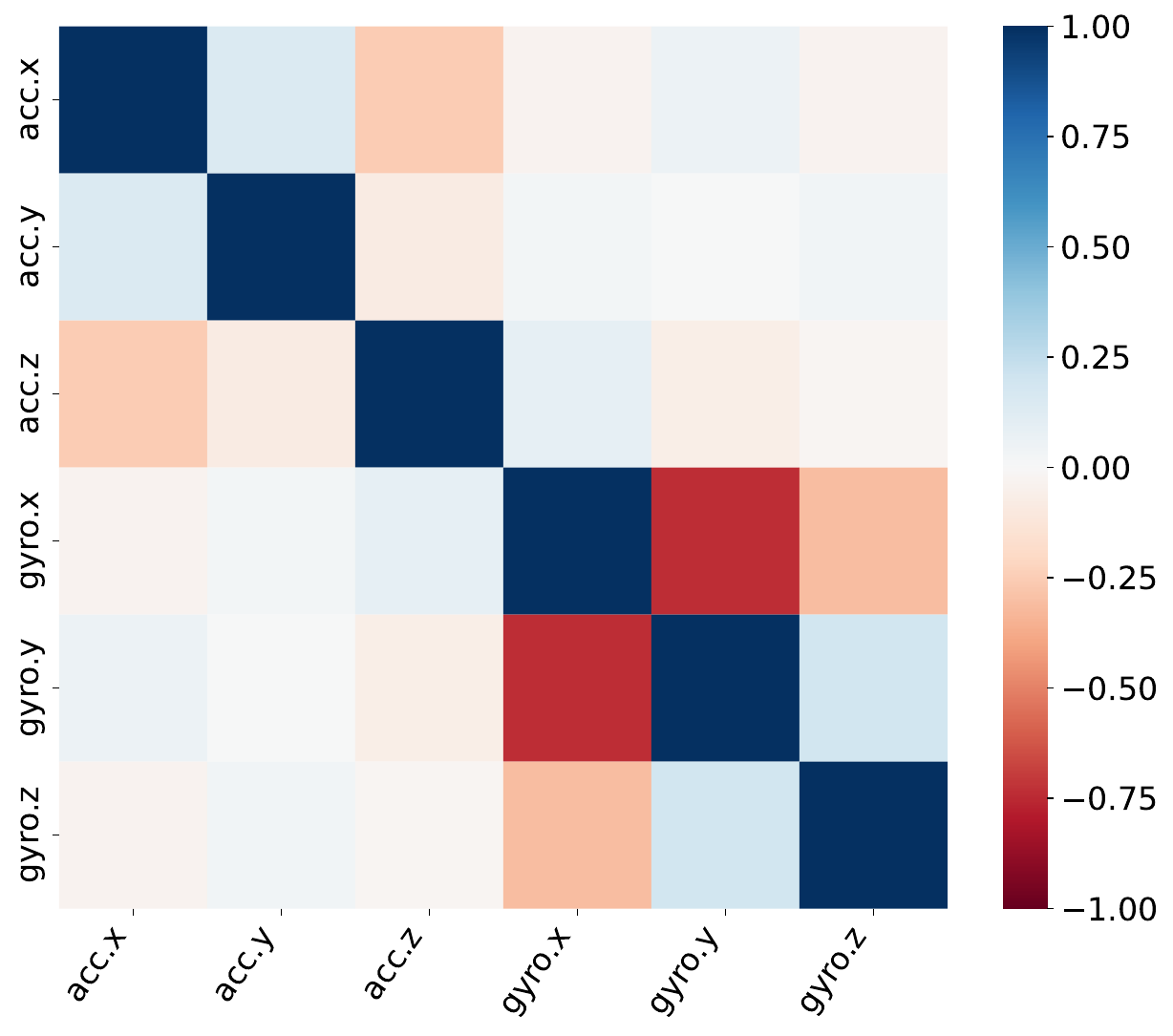}
        \caption{UCI-HAR}
        \label{subfig:acx}
    \end{subfigure}%
    \begin{subfigure}[t]{0.5\textwidth}
        \centering
        \includegraphics[trim={0 0 3.3cm 0},clip,width=\textwidth]{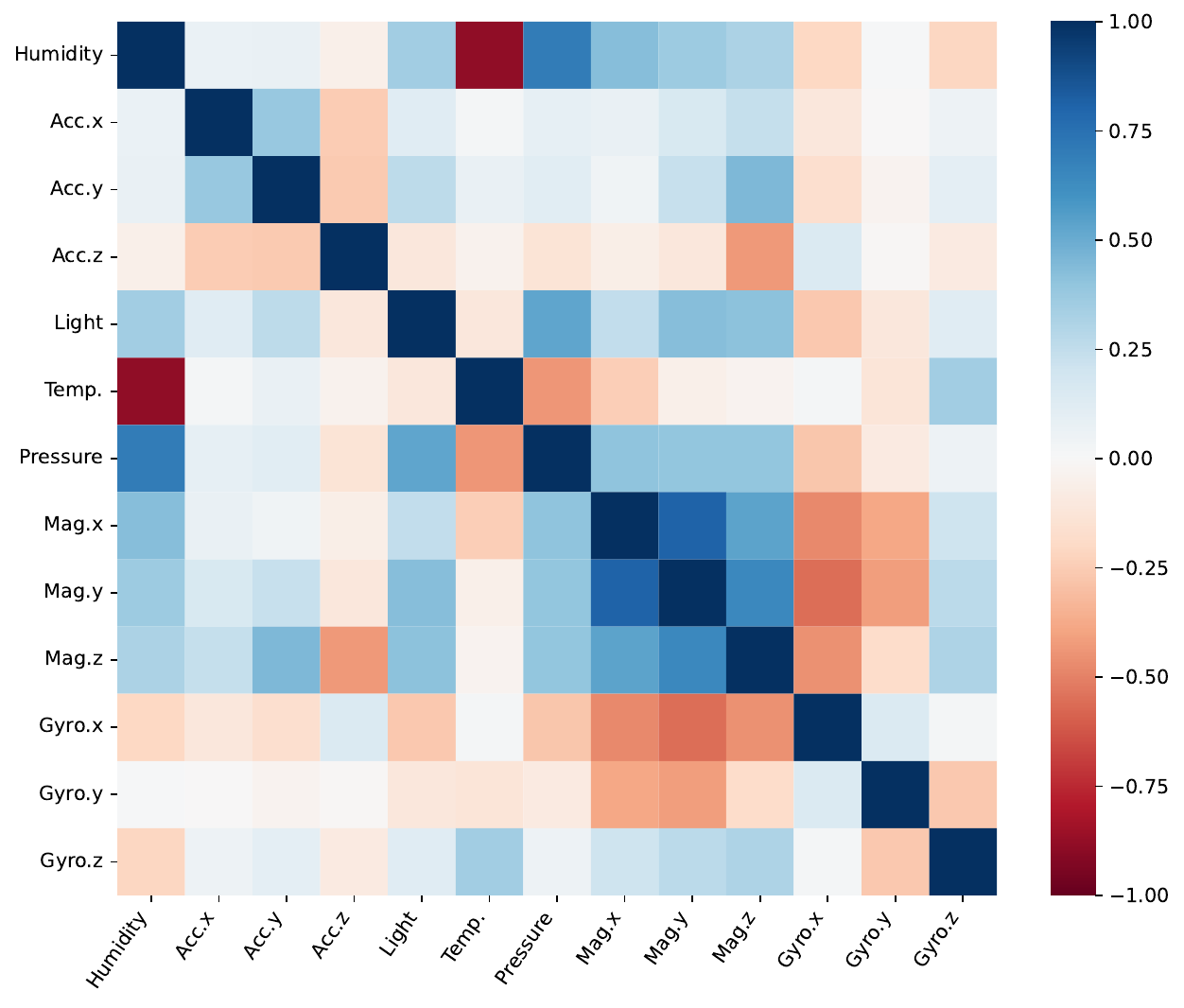}
        \caption{PerilZIS}
    \end{subfigure}%

    \begin{subfigure}[t]{0.45\textwidth}
        \centering
        \includegraphics[trim={0 0 3.3cm 0},clip,width=\textwidth]{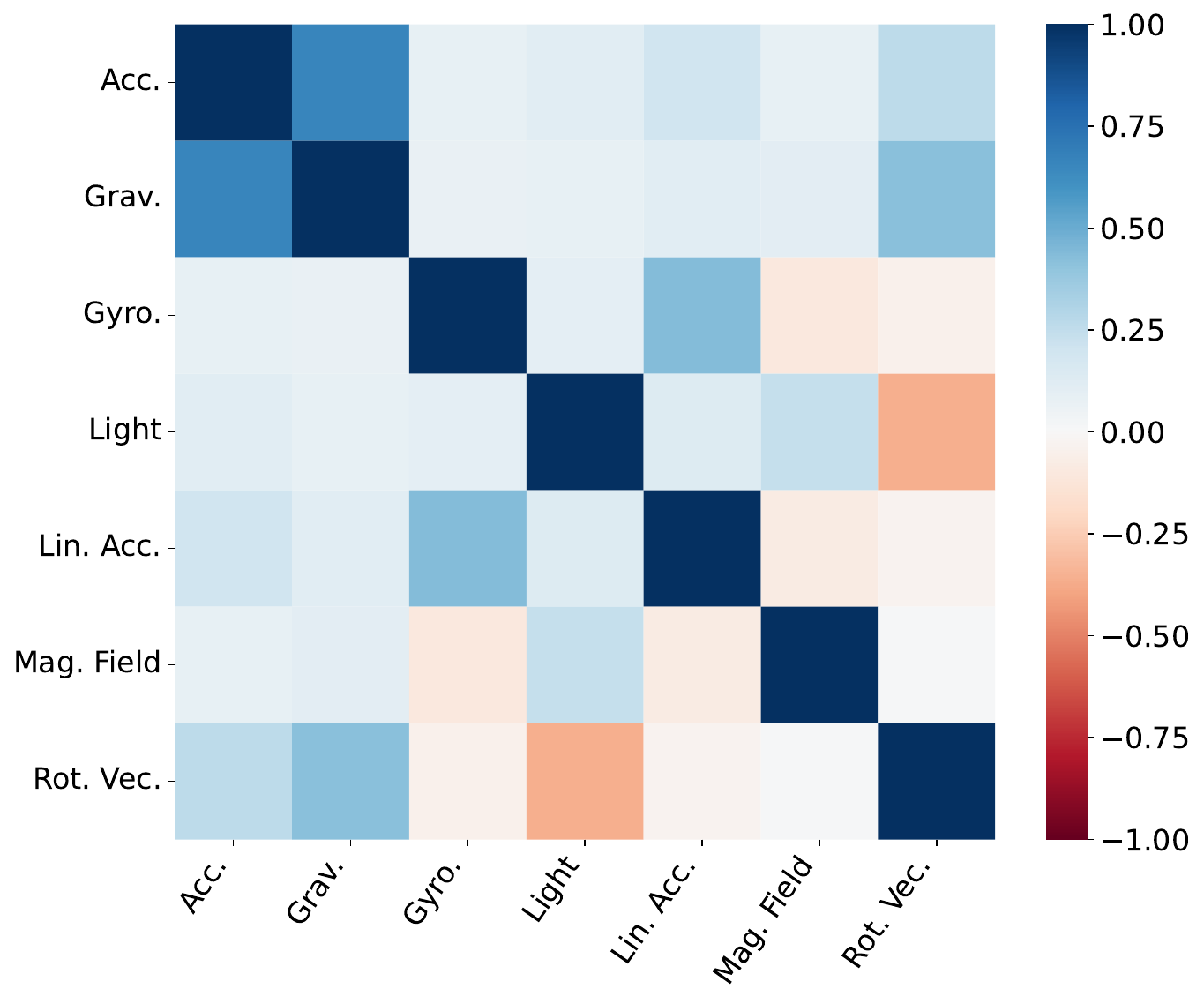}
        \caption{Relay}
    \end{subfigure}%
    \begin{subfigure}[t]{0.5\textwidth}
        \centering
        \includegraphics[trim={0 0 3.3cm 0},clip,width=\textwidth]{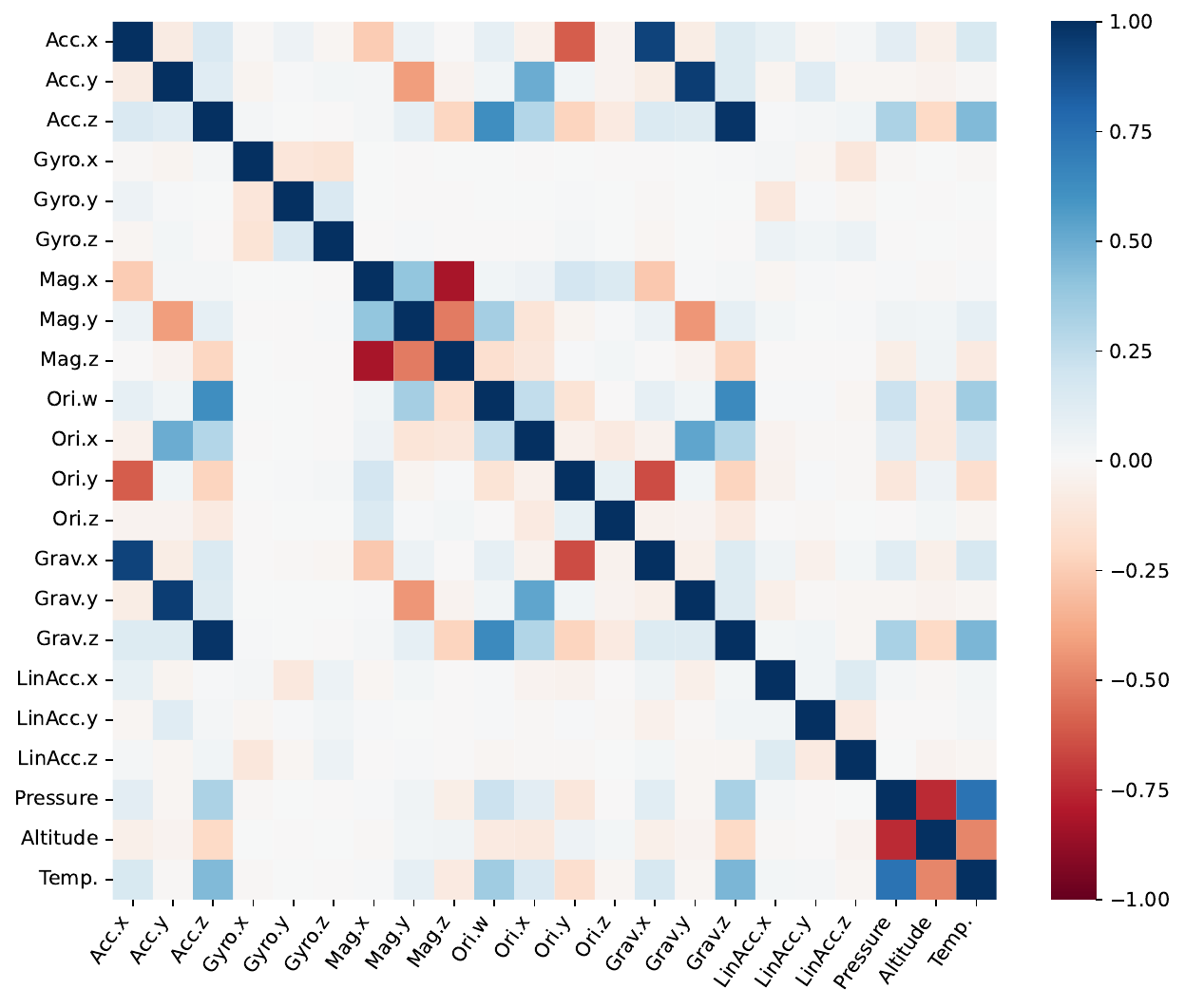}
        \caption{SHL}
    \end{subfigure}%
    \caption{Pairwise sensor correlation matrices for each dataset.}
    \label{fig:correlation-matrices}
\end{figure*}

Figure~\ref{fig:correlation-matrices} provides a quantitative view of inter-sensor redundancy through Pearson correlation matrices for each dataset. Strong correlations, with $|\rho|\approx 1$, indicate high linear dependence; such redundancy is a major contributor to fragility in the context of pairwise combinations. Employing correlated sensors adds little new unpredictability, which invalidates the simplistic additive model of entropy and can provide diminishing returns. Figure~\ref{fig:mi-matrices} (Appendix A) presents mutual information heatmaps. These results reveal dependencies between sensor pairs, confirming that redundancy extends beyond what Pearson correlation alone suggests.  Consequently, while modalities may appear as offering some level of unpredictability individually, the results suggest that overall \emph{joint} entropy can be sharply constrained compared to what would be expected if they were treated as independent variables.

\subsubsection{Complexity Challenges}
\label{sec:complexity-challenges}

Computing the exact joint probability distribution and joint entropy of multiple sensors quickly becomes infeasible. Consider the following: let each of the $n$ sensor modalities be discretised into $b_i$ bins, then the joint distribution has $\prod_{i=1}^{n} b_i$ distinct states. This produces an enormous state space as $n$ and $b_i$ grow to the values considered in this paper.  Applying the FD rule typically results in thousands of bins per modality, causing the number of joint bins to explode combinatorially.   In addition, even before enumerating states, \emph{selecting which sensors to combine} can itself involve $2^n - (n+1)$ subsets, skipping single-sensor subsets and the empty set. Preliminary experiments confirmed joint entropies could be computed directly for $n \leq 3$ modalities with a maximum of 1250 bins and fewer than 150K total samples from the Relay dataset. Reducing bin sizes can help, but this risks oversimplifying the distribution and artificially deflating entropy estimates (discussed in \S\ref{sec:preproc}). In our tests, we found that reducing bin counts reduced single-sensor entropy estimates by approximately 2--3 bits on average compared to those reported in Table~\ref{tab:single-sensor-results}.   We therefore adopt an alternative strategy that trades some accuracy for tractability.

\subsubsection{Chow-Liu Approximation}

To handle these scaling issues, we adopt \emph{Chow--Liu trees}~\cite{chow1968approximating} to approximate a high-dimensional joint distribution using a maximum-weight spanning tree, $\pi$, over the different sensor modalities. Each edge is weighted by the mutual information of the connected variables, ensuring the tree structure captures the dominant pairwise dependencies. This approach minimises the Kullback--Leibler divergence (Def.~\ref{sec:defs}) between the true multivariate distribution and the tree-based approximation as follows:

\begin{equation}
    p_{\pi}(x_1,\dots,x_n) =  
    p(x_r)\prod_{i\neq r}p(x_i\; |\; x_{\pi(x)})
\end{equation}

Where $\pi(i)$ denotes the parent of $X_i$ in the tree, and $r$ is the tree's root node. Chow--Liu trees are acyclic, singly connected structures: each node has at most one parent where one can traverse the tree to accumulate probabilities between pairwise dependencies. The tree captures dominant pairwise dependencies and significantly reduces computation time compared to na\"{i}ve enumeration of the full joint measurement space.  The use of Chow-Liu trees was proposed by Buller and Kaufer~\cite{buller2016estimating} for estimating the entropy of multivariate data sources where the range of possible values is high, as is the case with sensor outputs considered in this work. In our Python implementation, we use the pgmpy~\cite{ankan2024pgmpy} library's TreeSearch module to this end; practically, for each sensor subset, we:

\begin{enumerate}
    \item Discretise each sensor's readings via Freedman--Diaconis binning.
    \item Build a Chow--Liu tree from pairwise mutual information of sensors.
    \item Traverse the tree to estimate max ($H_0$), Shannon ($H_1$), collision ($H_2$), and min-entropy ($H_{\infty}$) without enumerating the full state space.
\end{enumerate}

\begin{table}
\centering
\caption{Average entropy results of Chow-Liu (CL) tree approximation against direct computation across all small sensor subsets ($n \in \{2,3\}$), giving the mean absolute error (MAE), averaged across all subsets, and the \% mean relative error.}
\label{tab:chowliu_validation_n23}
\small 
\resizebox{\linewidth}{!}{%
\begin{tabular}{l c cc cc cc cc cc}
\toprule
& & \multicolumn{2}{c}{\textbf{$H_0$}} & \multicolumn{2}{c}{\textbf{$H_1$}} & \multicolumn{2}{c}{\textbf{$H_2$}} & \multicolumn{2}{c}{\textbf{$H_\infty$}} & & \\
\cmidrule(lr){3-4} \cmidrule(lr){5-6} \cmidrule(lr){7-8} \cmidrule(lr){9-10}
\textbf{Dataset} & $n$ & {\textbf{Direct}} & {\textbf{CL}} & {\textbf{Direct}} & {\textbf{CL}} & {\textbf{Direct}} & {\textbf{CL}} & {\textbf{Direct}} & {\textbf{CL}} & {\textbf{MAE}} & {\textbf{\%}} \\
\midrule
\addlinespace 
UCI-HAR & 2 & 17.60 & 16.92 & 12.08 & 12.58 & 10.85 & 10.33 & 6.01 & 6.20 & 0.473 & 4.1 \\
        & 3 & 23.93 & 25.46 & 19.35 & 18.25 & 14.03 & 14.77 & 9.40 & 8.95 & 0.955 & 5.7 \\
\midrule
\addlinespace
Relay   & 2 & 15.36 & 15.54 & 9.72 & 10.12 & 8.27 & 7.95 & 4.64 & 4.88 & 0.285 & 3.0 \\
        & 3 & 21.47 & 22.84 & 16.84 & 15.89 & 13.69 & 13.04 & 7.41 & 7.96 & 0.875 & 5.5 \\
\midrule
\addlinespace
PerilZIS& 2 & 20.33 & 22.18 & 14.77 & 14.00 & 11.40 & 12.18 & 7.21 & 8.52 & 1.013 & 7.6 \\
        & 3 & 30.82 & 32.55 & 21.28 & 20.08 & 16.66 & 17.65 & 11.46 & 12.06 & 1.258 & 6.2 \\
\midrule
\addlinespace
SHL     & 2 & 24.63 & 26.25 & 18.76 & 18.20 & 14.17 & 14.28 & 9.50 & 8.92 & 0.693 & 4.1 \\
        & 3 & 36.71 & 39.05 & 28.44 & 26.83 & 22.46 & 21.14 & 12.44 & 13.25 & 1.488 & 4.9 \\
\bottomrule
\end{tabular}
}
\end{table}

We validated the approach against a ground-truth derived by computing the joint entropy through exhaustive enumeration. As this direct method is only computationally feasible for a small number of modalities, we performed the comparison on all two- and three-sensor subsets from each dataset. The results of this validation are shown in Table~\ref{tab:chowliu_validation_n23}. Across all datasets, the estimates from Chow-Liu trees closely track direct ground-truth calculations in these low-dimensional cases. The relative error is consistently low (3--8\%), supporting its use within higher-dimensional combinations where direct computation is infeasible.\footnote{We note the caveat that efficiently estimating the entropy of high-dimensionality sources remains an ongoing research challenge~\cite{buller2016estimating}. We must emphasise that our results are indicative estimations; quantifying errors beyond \(n\geq4\) is an open problem.} Following this, we proceeded to estimate the entropy for all modalities in each dataset. Our developed framework evaluates the joint entropy over all sensor combinations; the powerset of the sensor set is generated and processed in parallel using Python's multiprocessing module. Processing all datasets took approximately 22 hours on our workstation with an Intel i7-6700K (8M cache, 4.20 GHz) CPU and 32 GB RAM on Ubuntu 24.04.

\subsubsection{Results}
\label{subsubsec:results}

Tables~\ref{tab:top10-uci-har}--\ref{tab:top10-shl} report the top 10 performing multi-sensor combinations ranked by min-entropy for each dataset. Combining \emph{all} sensors yields the highest $H_0$ (max-entropy) and often increases Shannon and collision entropy. However, min-entropy ($H_{\infty}$) remains stubbornly low;  the complete set of sensors in SHL, for instance, surpasses 80 bits of $H_1$, but saturates at only 21\,bits of $H_{\infty}$.  Interestingly, we find that \emph{omitting} certain correlated sensors sometimes barely reduces min-entropy at all. For the Relay dataset in Table~\ref{tab:top10-relay}, the combination (Acc., Gyro., Light, Lin.~Acc., Mag., Rot.~Vec.) achieves \(H_{\infty} = 7.859\) bits, only slightly below the \(8.092\) bits when using every sensor. Parallel findings arise in the PerilZIS and SHL datasets, where omitting a small number of sensors from the ``All sensors'' set has negligible impact on \(H_{\infty}\). This pattern appears across datasets: additional modalities raise $H_0$ and $H_1$ but barely move $H_{\infty}$.

\begin{figure*}[t!]
    \centering
    \begin{subfigure}[t]{0.5\textwidth}
        \centering
        \includegraphics[width=\textwidth]{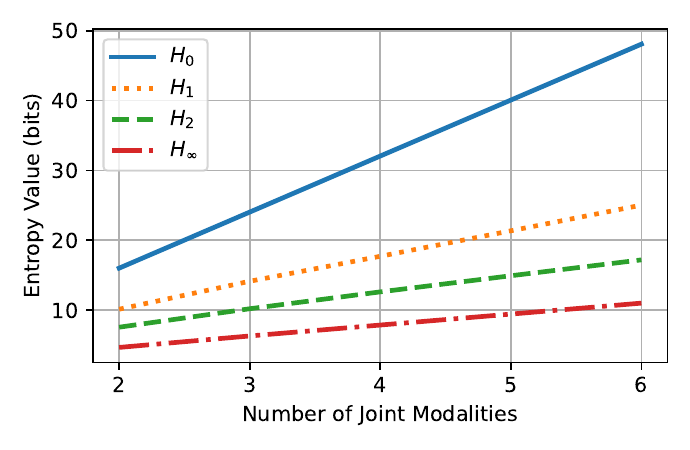}
        \caption{UCI-HAR}
        \label{subfig:acx}
    \end{subfigure}%
    \begin{subfigure}[t]{0.5\textwidth}
        \centering
        \includegraphics[width=\textwidth]{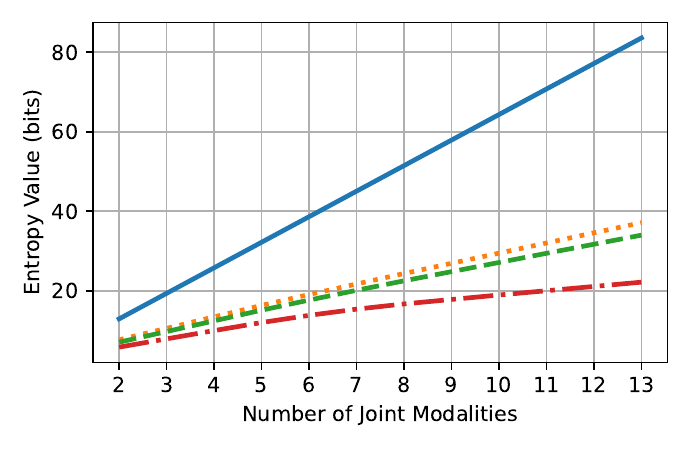}
        \caption{PerilZIS}
    \end{subfigure}%

    \begin{subfigure}[t]{0.5\textwidth}
        \centering
        \includegraphics[width=\textwidth]{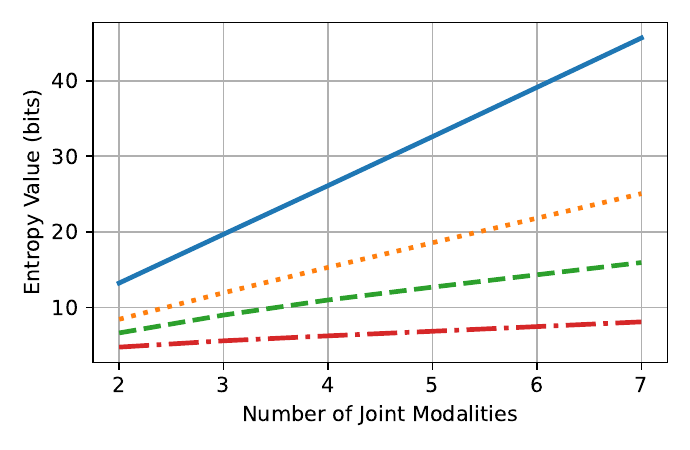}
        \caption{Relay}
    \end{subfigure}%
    \begin{subfigure}[t]{0.5\textwidth}
        \centering
        \includegraphics[width=\textwidth]{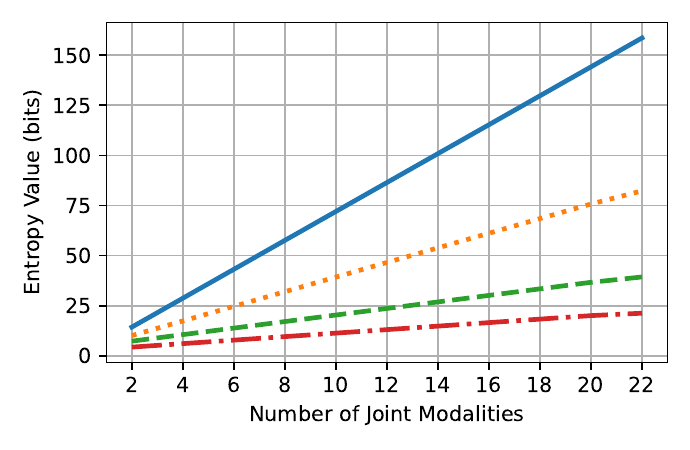}
        \caption{SHL}
    \end{subfigure}%
    \caption{Average joint entropy vs.\ number of combined sensors modalities.}
    \label{fig:h-graphs}
\end{figure*}

We plot the number of combined modalities for each dataset in Figure~\ref{fig:h-graphs}, showing the average (mean) joint entropy for different entropy measures. The data illustrates that, while $H$ values are tightly coupled initially, they diverge significantly as the number of joint modalities increases. All datasets demonstrated a trend of worst-case $H_{\infty}$ remaining significantly lower and growing only marginally with respect to other entropy values. For SHL and PerilZIS, for example, $H_{\infty}$ increases by approximately only 1--2 bits per modality, with $H_{0}$--$H_2$ increasing far more significantly ($\Delta H_{0} \approx$ 10 bits and $\Delta H_{1} \approx$ 5 bits for every added modality). In Tables~\ref{tab:top10-uci-har}--\ref{tab:top10-shl}, we give the differences of $H_1$ and $H_{\infty}$ for the top combinations as \(H_1 - H_{\infty}\), quantifying the sharp decreases between Shannon and min-entropy as the number of combined modalities increases. We also use \# to denote the number of modalities used in each row. 

These results confirm that many sensors contribute little beyond redundancy when combined, revealing a fundamental limitation of multi-modal data in practice. While joint signals raise the apparent entropy, correlations between modalities prevent a corresponding rise in min-entropy; large ensembles offer no substantial advantage over smaller subsets of sensors.  This underscores why min-entropy is the relevant metric for worst-case guarantees in modern standards~\cite{bsi2024ais31,turan2018recommendation}. To our knowledge, the extent of \emph{entropy collapse} under correlated mobile modalities has not been elicited at this scale. Full combination results are available in our repository (see Footnote~\ref{fn:url}).

\begin{table}[ht!]
\centering
\caption{Top 10 best-performing sensor combinations (UCI-HAR; in bits).}
\resizebox{0.7\linewidth}{!}{%
\begin{tabular}{@{}r|c|ccc|c|c@{}}
\toprule
\textbf{Modality} & \textbf{\#} & \( H_{0} \) & \( H_{1} \) & \( H_{2} \) & \( H_{\infty} \) & \textbf{\(H_1 - H_{\infty}\)} \\ 
\midrule
All sensors & 6 & 48.061 & 25.089 & 17.116 & 11.008 & 14.081 \\
\midrule
(Acc.\{x,y,z\}, Gyro.\{y,z\}) & 5 & 39.403 & 22.835 & 15.999 & 10.113 & 12.722 \\
(Acc.\{x,y,z\}, Gyro.\{x,y\}) & 5 & 39.886 & 21.456 & 15.332 & 9.900 & 11.556 \\
(Acc.\{x,y,z\}, Gyro.\{x,z\}) & 5 & 40.222 & 21.470 & 15.040 & 9.411 & 12.059 \\
(Acc.\{y,z\}, Gyro.\{x,y,z\}) & 5 & 40.228 & 21.306 & 14.698 & 9.274 & 12.032 \\
(Acc.\{x,z\}, Gyro.\{x,y,z\}) & 5 & 40.293 & 21.260 & 14.575 & 9.154 & 12.106 \\
(Acc.\{x,y,z\}, Gyro.y) & 4 & 31.228 & 18.804 & 13.893 & 9.065 & 9.739 \\
(Acc.\{x,y\}, Gyro.\{x,y,z\}) & 5 & 40.273 & 21.155 & 14.208 & 8.721 & 12.434 \\
(Acc.\{x,y,z\}, Gyro.z) & 4 & 31.564 & 18.775 & 13.833 & 8.682 & 10.093 \\
(Acc.\{y,z\}, Gyro.\{y,z\}) & 4 & 31.570 & 18.583 & 13.400 & 8.386 & 10.197 \\
\bottomrule
\end{tabular}
}
\label{tab:top10-uci-har}
\end{table}

\begin{table}[ht!]
\centering
\caption{Top 10 best-performing sensor combinations (Relay; in bits).}
\resizebox{0.9\linewidth}{!}{%
\begin{tabular}{@{}r|c|ccc|c|c@{}}
\toprule
\textbf{Modality} & \textbf{\#} & \( H_{0} \) & \( H_{1} \) & \( H_{2} \) & \( H_{\infty} \) & \textbf{\(H_1 - H_{\infty}\)} \\ 
\midrule
All sensors & 7 & 45.794 & 25.036 & 15.725 & 8.092 & 16.944 \\
\midrule
(Acc., Gyro., Light, Lin. Acc., Mag., Rot. Vec.) & 6 & 44.795 & 24.963 & 15.334 & 7.859 & 17.104 \\
(Acc., Grav., Light, Lin. Acc., Mag., Rot. Vec.) & 6 & 38.385 & 21.590 & 14.783 & 7.702 & 13.888 \\
(Acc., Grav., Gyro., Light, Mag., Rot. Vec.) & 6 & 37.026 & 21.001 & 14.553 & 7.702 & 13.299 \\
(Grav., Gyro., Light, Lin. Acc., Mag., Rot. Vec.) & 6 & 36.092 & 20.794 & 14.450 & 7.673 & 13.121 \\
(Acc., Light, Lin. Acc., Mag., Rot. Vec.) & 5 & 37.385 & 21.517 & 14.428 & 7.469 & 14.048 \\
(Acc., Gyro., Light, Mag., Rot. Vec.) & 5 & 36.026 & 20.924 & 14.270 & 7.465 & 13.459 \\
(Gyro., Light, Lin. Acc., Mag., Rot. Vec.) & 5 & 35.092 & 20.720 & 14.147 & 7.439 & 13.281 \\
(Acc., Grav., Gyro., Light, Lin. Acc., Mag.) & 6 & 41.094 & 22.794 & 14.437 & 7.338 & 15.456 \\
(Acc., Grav., Light, Mag., Rot. Vec.) & 5 & 29.617 & 17.555 & 13.244 & 7.312 & 10.243 \\
\bottomrule
\end{tabular}
}
\label{tab:top10-relay}
\end{table}

\begin{table}[ht!]
\centering
\caption{Top 10 best-performing sensor combinations (PerilZIS; in bits).}
\resizebox{\linewidth}{!}{%
\begin{tabular}{@{}r|c|ccc|c|c@{}}
\toprule
\textbf{Modalities} & \textbf{\#} & \( H_{0} \) & \( H_{1} \) & \( H_{2} \) & \( H_{\infty} \) & \textbf{\(H_1 - H_{\infty}\)} \\ 
\midrule
All sensors & 13 & 83.662 & 36.998 & 33.682 & 23.926 & 13.072 \\
\midrule
(Acc.\{x,y,z\}, Light, Temp., Pres., Mag.\{x,y,z\}, Gyro.\{x,y,z\}) & 12 & 76.907 & 34.258 & 31.510 & 23.835 & 10.423 \\
(Acc.\{y,z\}, Light, Temp., Pres., Mag.\{x,y,z\}, Gyro.\{x,y,z\}) & 11 & 72.737 & 34.246 & 31.509 & 23.835 & 10.411 \\
(Acc.\{x,z\}, Light, Temp., Pres., Mag.\{x,y,z\}, Gyro.\{x,y,z\}) & 11 & 72.820 & 34.243 & 31.508 & 23.829 & 10.414 \\
(Acc.\{x,y\}, Light, Temp., Pres., Mag.\{x,y,z\}, Gyro.\{x,y,z\}) & 11 & 72.737 & 34.229 & 31.501 & 22.945 & 11.284 \\
(Acc.\{x,y,z\}, Light, Temp., Mag.\{x,y,z\}, Gyro.\{x,y,z\}) & 11 & 71.384 & 32.586 & 30.190 & 22.945 & 9.641 \\
(Acc.y, Light, Temp., Press., Mag.\{x,y,z\}, Gyro.\{x,y,z\}) & 10 & 68.568 & 34.218 & 31.500 & 22.945 & 11.273 \\
(Acc.x, Light, Temp., Pres., Mag.\{x,y,z\}, Gyro.\{x,y,z\}) & 10 & 68.650 & 34.215 & 31.499 & 22.945 & 11.270 \\
(Acc.\{y,z\}, Light, Temp., Mag.\{x,y,z\}, Gyro.\{x,y,z\}) & 10 & 67.214 & 32.575 & 30.189 & 22.945 & 9.630 \\
(Acc.\{x,y,z\}, Light, Temp., Pres., Mag.\{x,z\}, Gyro.\{x,y,z\}, Hum.) & 12 & 76.618 & 33.816 & 31.035 & 21.915 & 11.901 \\
\bottomrule
\end{tabular}
}
\label{tab:top10-peril}
\end{table}

\begin{table}[ht!]
\centering
\caption{Top 10 best-performing sensor combinations (SHL; in bits).}
\resizebox{\linewidth}{!}{%
\begin{tabular}{@{}R{10cm}|c|ccc|c|c@{}}
\toprule
\textbf{Modality} & \textbf{\#} & \( H_{0} \) & \( H_{1} \) & \( H_{2} \) & \( H_{\infty} \) & \textbf{\(H_1 - H_{\infty}\)} \\ 
\midrule
All sensors & 22 & 158.601 & 82.301 & 39.320 & 21.289 & 61.012 \\
\midrule
(Acc.\{x,y,z\}, Gyro.\{x,y,z\}, Mag.\{x,y,z\}, Ori.\{w,x,y,z\}, Grav.\{x,y,z\}, LinAcc.\{y,z\}, Pres., Alt., Temp.) & 21 & 148.995 & 78.624 & 39.276 & 21.289 & 57.335 \\
(Acc.\{x,y,z\}, Gyro.\{x,y,z\}, Mag.\{x,y,z\}, Ori.\{w,x,y,z\}, Grav.\{x,y,z\}, LinAcc.\{x,z\}, Pres., Alt., Temp.) & 21 & 148.759 & 78.477 & 39.272 & 21.289 & 57.188 \\
(Acc.\{x,y,z\}, Gyro.\{x,z\}, Mag.\{x,y,z\}, Ori.\{w,x,y,z\}, Grav.\{x,y,z\}, LinAcc.\{x,y,z\}, Pres., Alt., Temp.) & 21 & 148.587 & 78.380 & 39.249 & 21.289 & 57.091 \\
(Acc.\{x,y,z\}, Gyro.\{x,y,z\}, Mag.\{x,y,z\}, Ori.\{w,x,y,z\}, Grav.\{x,y,z\}, LinAcc.\{x,y\}, Pres., Alt., Temp.) & 21 & 148.952 & 78.135 & 39.235 & 21.289 & 56.846 \\
(Acc.\{x,y,z\}, Gyro.\{x,y,z\}, Mag.\{x,y,z\}, Ori.\{w,x,y,z\}, Grav.\{x,y,z\}, LinAcc.\{z\}, Pres., Alt., Temp.) & 20 & 139.153 & 74.800 & 39.175 & 21.289 & 53.511 \\
(Acc.\{x,y,z\}, Gyro.\{x,z\}, Mag.\{x,y,z\}, Ori.\{w,x,y,z\}, Grav.\{x,y,z\}, LinAcc.\{y,z\}, Pres., Alt., Temp.) & 20 & 138.981 & 74.703 & 39.128 & 21.289 & 53.414 \\
(Acc.\{x,y,z\}, Gyro.\{x,z\}, Mag.\{x,y,z\}, Ori.\{w,x,y,z\}, Grav.\{x,y,z\}, LinAcc.\{x,z\}, Pres., Alt., Temp.) & 20 & 138.745 & 74.556 & 39.117 & 21.289 & 53.267 \\
(Acc.\{x,y,z\}, Gyro.\{x,y,z\}, Mag.\{x,y,z\}, Ori.\{w,x,y,z\}, Grav.\{x,y,z\}, LinAcc.\{y\}, Pres., Alt., Temp.) & 20 & 139.346 & 74.458 & 39.100 & 21.289 & 53.169 \\
(Acc.\{x,y,z\}, Gyro.\{x,y,z\}, Mag.\{x,y,z\}, Ori.\{w,x,y,z\}, Grav.\{x,y,z\}, LinAcc.\{x\}, Pres., Alt., Temp.) & 20 & 139.109 & 74.311 & 39.087 & 21.289 & 53.022 \\
\bottomrule
\end{tabular}
}
\label{tab:top10-shl}
\end{table}

\subsubsection{Binning Sensitivity Analysis}
\label{sec:bin-sensitivity}

\begin{figure*}[t!]
    \centering
    \begin{subfigure}[t]{0.49\textwidth}
        \centering
        \includegraphics[width=\textwidth]{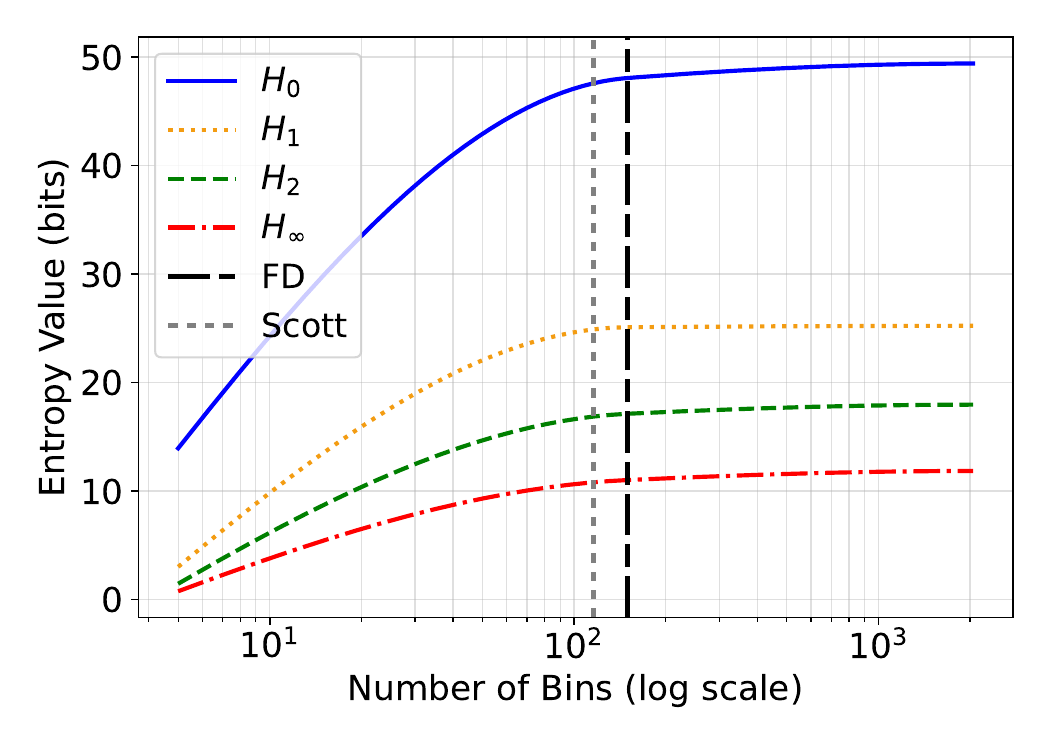}
        \caption{UCI-HAR}
        \label{subfig:uciharbin}
    \end{subfigure}%
    \begin{subfigure}[t]{0.49\textwidth}
        \centering
        \includegraphics[width=\textwidth]{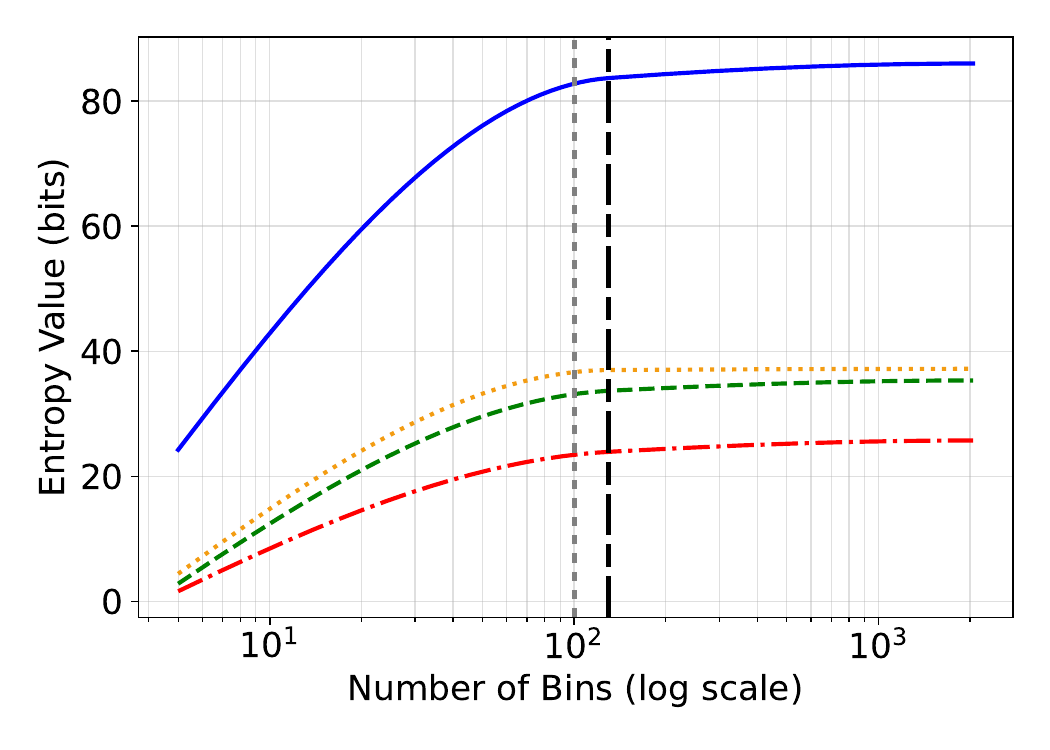}
        \caption{PerilZIS}
    \end{subfigure}%

    \begin{subfigure}[t]{0.49\textwidth}
        \centering
        \includegraphics[width=\textwidth]{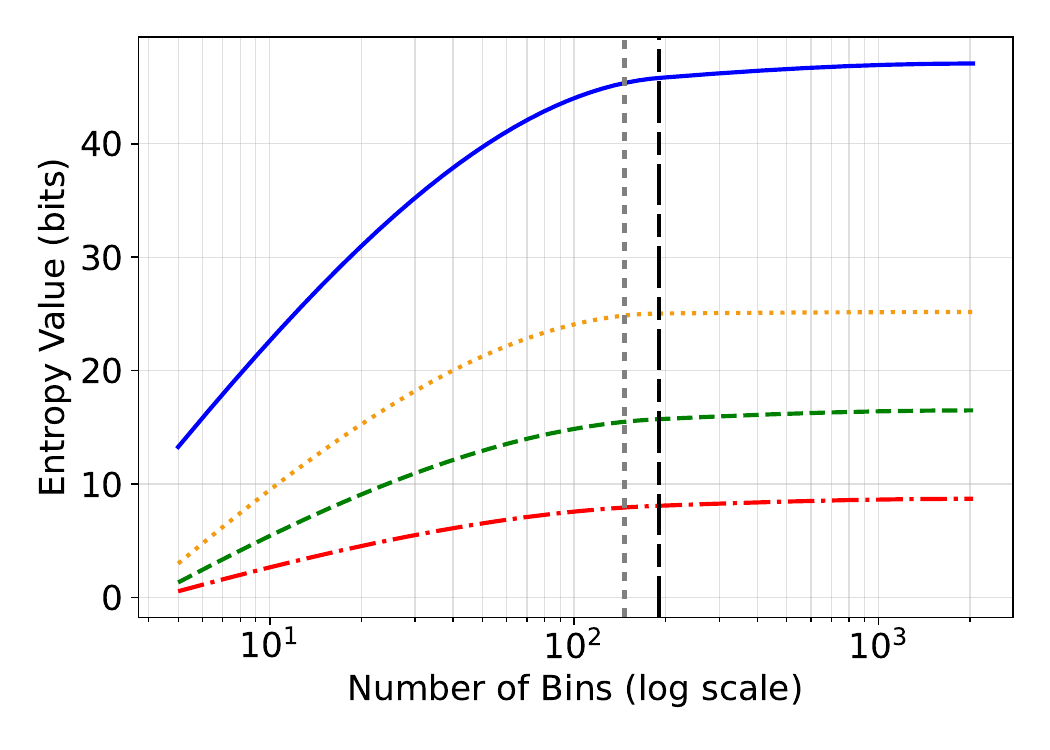}
        \caption{Relay}
    \end{subfigure}%
    \begin{subfigure}[t]{0.49\textwidth}
        \centering
        \includegraphics[width=\textwidth]{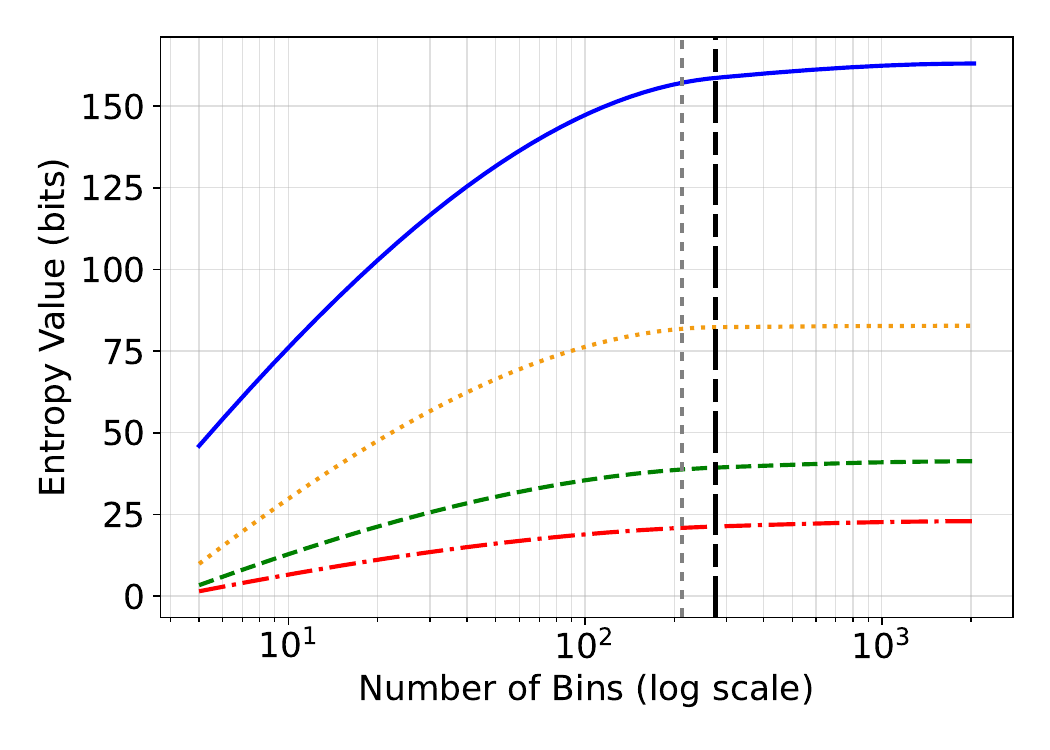}
        \caption{SHL}
    \end{subfigure}%
    \caption{Bin sensitivity analysis showing how entropy values vary by average bin width for the best-performing sensor combinations for each data set. Freedman--Diaconis (FD) and Scott's rule values are also shown.}
    \label{fig:bins-graphs}
\end{figure*}

As we discussed in \S\ref{sec:preproc} and \S\ref{sec:complexity-challenges}, bin width is a key consideration in estimating entropy accurately and efficiently. Too few bins cause oversmoothing, erasing distributional features and deflating entropy. At the same time, too many bins induces sparsity and overfitting, inflating entropy by modelling noise rather than the source. We performed a sensitivity analysis on the best-performing sensor combination (``All sensors'') for each of our four datasets. We re-calculated the joint entropy while varying the number of bins for each modality, starting from 5 bins and increasing up to 2048. Beyond that, the computational cost of processing the large state space became prohibitive. The results of this analysis are presented in Figure~\ref{fig:bins-graphs}. For completeness, we also give an alternative binning strategy---Scott's rule ($3.5\sigma \cdot n^{-1/3}$)---as a comparator for our FD decision rule, which we report as a stability check.

The graphs show a consistent trend across all datasets. For a low number of bins, all entropy measures are severely underestimated as expected. As the number of bins increases, the entropy rises sharply before converging towards a stable plateau that approximates a logarithmic shape. (An exception is the min-entropy of SHL, which converges after $\approx$100 bins). Crucially, FD-selected bin counts fall within the stable region, supporting FD as a balanced choice between fidelity and tractability. 

\section{Evaluation} \label{sec:security-evaluation}

We now discuss the key takeaways, relate min-entropy to concrete guessing cost, discuss mitigations, and give recommendations for future research in the area.

\subsection{Discussion}

Prior studies have directly claimed or implicitly assumed that mobile sensors are suitable data sources for security-critical applications. To this end, work has relied on model evaluation metrics as a proxy for evaluating such claims~\cite{mehrnezhad2015tap,gurulian2018good,markantonakis2024using,shrestha2014drone}, with a smaller subset using more established entropy metrics~\cite{wu2024t2pair,li2020t2pair}. Separate entropy analyses of sensors have demonstrated that \emph{individual} sensors actually confer very little entropy, particularly under worst-case assumptions~\cite{lv2020analysis,hennebert2013entropy,voris2011accelerometers,krhovjak2007sources}. Our results both challenge and extend these claims.

\textbf{Single sensors yield low entropy.} Building on prior work, we find that sensor data has stubbornly low min-entropy (e.g.\ some single-sensor readings yield only 1--3 bits) and remains significantly below Shannon entropy.  This exposes a major gap between average- and worst-case unpredictability across 25 widely used sensors---a critical distinction that existing schemes and evaluations do not capture at this scale.

\textbf{Combining multiple sensors offers modest gains with respect to worst-case min-entropy.} While using multi-modal sensors typically raises best- and average-case entropies, our experiments reveal that gains in worst-case min-entropy are far smaller than one might expect. In many cases, $H_{\infty}$ plateaus, reflecting the dominance of a handful of high-probability joint outcomes. Even the most complex multi-modal combinations in our datasets yield only modest worst-case entropy---between 8--24 bits in the best cases---which is well below the levels required by modern security standards. This also aligns with prior work showing that Shannon entropy can significantly overstate security~\cite{arikan2002inequality,cachin1997entropy,turan2018recommendation,bsi2024ais31}. Hence, while mobile sensors can \emph{augment} other authentication or key-generation processes, they do not suffice by themselves. Using multiple sensors may still improve usability and model accuracy, e.g.\ for coarse co-location checks, but they should not be treated as secure, unpredictable sources.

\subsection{Translating Entropy Bounds Into Guessing Attacks}

We now sketch how low-entropy sensor data translates into concrete attack cost. Let $Z$ be a random variable over the discretised joint sensor measurement space $\mathcal{Z}$ used by a scheme, and let $H_{\infty}(Z)$ denote its min-entropy. If an adversary enumerates candidates $z \in \mathcal{Z}$ in order of decreasing probability $\Pr[Z = z]$, then after $q$ guesses the success probability is bounded by
\begin{equation}
\Pr[\text{success within } q] \;\le\; q \cdot 2^{-H_\infty(Z)}.
\label{eq:qbound}
\end{equation}

The expected number of guesses is $ E[G]\approx2^{H_\infty(Z)-1}$~\cite{cachin1997entropy} in order to identify the correct sensor vector.

Most systems accept \emph{similar}, not exact, matches through learned decision boundaries or fuzzy matching~\cite{mehrnezhad2015tap,gurulian2018good,shrestha2014drone,markantonakis2024using,alzubaidi2016authentication,mayrhofer2021adversary,abuhamad2020sensor}. Let $\mathsf{Auth} : \mathcal{Z} \to \{0,1\}$ be an abstract authentication function that accepts ($1$) or rejects ($0$) an input vector $z \in \mathcal{Z}$. Under such a similarity-based scheme, the adversary does not need to reproduce the exact enrolled sensor vector; it is sufficient to generate some $z'$ that lies within the acceptance region of a high-probability template. This means the effective success probability for each guess is higher than the bound in Eq.~\ref{eq:qbound} suggests. The best-performing combination in our study yielded $H_{\infty} \approx 24$ bits, corresponding to a search space of $2^{24}$ ($\approx$16.7 million) possibilities. An adversary may pre-compute the most likely vectors and submit them to $\mathrm{Auth}$, requiring only $2^{23}$ guesses on average. This is well within the capabilities of a moderately resourced adversary---taking seconds to milliseconds on a modern PC to enumerate---and is orders of magnitude weaker than the $2^{128}$ expected of standard cryptographic keys~\cite{nistsp80057}. Indeed, prior work has already demonstrated the feasibility of guessing biometric templates using this approach~\cite{mai2017guessability}. In the worst cases, such as single-sensor modalities where $H_{\infty} < 8$ bits, the search space is trivial for an attacker to exhaust, requiring $\approx 2^7$ guesses on average.  We shall now give more concrete indicative guesswork efforts in line with Cachin~\cite{cachin1997entropy}. Consider an adversary who has unlimited guess attempts against $\mathrm{Auth}$, which returns an immediate success/failure result. We assume the attacker submits $q$ queries, i.e.\ measurement(s), in descending order of their probability. The attacker effort can thus be characterised by the results given in Table~\ref{tab:attk}.  A primary defence here is to rate-limit the number of attempts; if each guess requires a live interaction, and the system enforces a limit of $r$ attempts per second, the expected time to success is scaled to $\approx 2^{H_{\infty}-1}/r$. Nevertheless, such limits linearly stretch attack time but do not change the fundamental feasibility of guessing when $H_{\infty}(Z)$ is in the low-to-mid tens of bits.

\begin{table}[t]
\centering
\footnotesize
\caption{Illustrating attacker effort versus min-entropy under optimal online guessing.}
\label{tab:attk}
\begin{tabular}{r r r r r r}
\toprule
$H_{\infty}$ & $E[G]$ (guesses) & Time $q{=}1$/s & $q{=}10$/s & $q{=}10^3$/s &  $q{=}10^6$/s \\
\midrule
8  & $128$            & $128$ s              & $12.8$ s            & $0.128$ s          & $0.128$ ms \\
12 & $2{,}048$        & $34.1$ min           & $3.41$ min          & $2.05$ s           & $2.05$ ms \\
16 & $32{,}768$       & $9.10$ h             & $54.6$ min          & $32.8$ s           & $3.28$ ms \\
20 & $524{,}288$      & $6.07$ d             & $14.6$ h            & $8.74$ min         & $0.524$ s \\
24 & $8.39\times 10^6$& $97.1$ d             & $9.71$ d            & $2.33$ h           & $8.39$ s \\
\bottomrule
\end{tabular}
\end{table}

\subsection{What can be done?}

Our results highlight a core tension in sensor-based security design. Proposals for pairing, co-location inference, or shared-secret agreement typically depend on the \emph{similarity} of sensor signals across devices. Yet the very regularities that enable similarity seemingly stem from \emph{low-entropy} structure. In effect, these systems trade cryptographic unpredictability for application-level utility. The open question is whether a scheme can ever achieve both meaningful entropy and the cross-device similarity it depends on. We close by offering several recommendations for researchers in this area.

\paragraph{Evaluate the threat of adversarial samples} A persistent feature in the literature is the lack of rigorous analysis of \emph{min-entropy} (and induced guesswork) under realistic acceptance regions and attacker priors~\cite{fomichev2019perils,mehrnezhad2015tap,shrestha2014drone,shrestha2018sensor,markantonakis2024using,mayrhofer2009shake,gurulian2017effectiveness,gurulian2018good}. We urge the community to consider the resilience of proposals to adversarial samples, particularly when using `fuzzy' matching methods such as supervised learning models, and to translate those estimates to concrete costs, such as expected guesses.

\paragraph{Quantify effects of deliberate interaction and induced phenomena}     User actions such as prescribed shakes or gestures, or externally induced signals, can raise entropy by several bits~\cite{voris2011accelerometers,lv2020analysis}. For example, altering the device's environment to produce distinctive electromagnetic signatures may be detectable by onboard sensors. A key gap is to quantify how much $\Delta H_\infty$ such interventions can actually deliver across modalities. Our expectation is that any gains remain modest, but this trade-off between added entropy and practical feasibility remains an under-explored area of research.

\paragraph{Decouple sensor similarity from unpredictability} Where applications need similarity evidence, such as for device authentication, one should treat that evidence as weak, predictable inputs. The derivation of shared secrets must come from a component designed for unpredictability; any sensor contribution should be considered, at most, as a small additive input to an entropy pool (see \cite{suciu2011unpredictable,wallace2016toward}). To improve the statistical quality of sensor-derived noise, one may apply cryptographic extractors (e.g.\ the Von Neumann extractor and newer schemes~\cite{von195113,trevisan2001extractors,barak2003true}) or decorrelation methods that attenuate dominant, deterministic trends. For example, principal component analysis (PCA) has been used elsewhere to isolate high-variance, predictable components---for example, gravitational bias in accelerometer axes---from low-variance residuals dominated by stochastic noise, which may then be discretised into candidate bitstrings (e.g.~\cite{liu2025post}). The data processing inequality is relevant here, where $H_{\infty}(f(X)) \leq H_{\infty}(X)$, for any deterministic transformation, $f$~\cite{vadhan2012pseudorandomness}.   That is, while extractors and transformations can reduce bias and correlation, they cannot increase intrinsic min-entropy; if sensors offer $k$ bits of $H_{\infty}$, post-processing can yield (at best) an unbiased string of length $\approx k$, and no more. Thus, such transformations are best interpreted as preconditioning or whitening steps that facilitate later modelling or extraction, but cannot increase the underlying min-entropy of the source itself.  Future work should focus on designing sensor-based schemes that maximise shared entropy between devices while preserving the practical utility of sensor data for their intended applications. %Extractors improve the quality of usable randomness when sufficient $H_{\infty}$ already exists; they do not increase the quantity in a way that raises brute-force cost. Future work in the area should focus on how entropy can be meaningfully increased while retaining application utility.%As such, transformations are best viewed as a decorrelation or preconditioning step, which simplify modelling, but cannot increase min-entropy beyond the source.

\subsection{Scope and Limitations}

While we analyse four large datasets, they do not fully capture contexts all possible contexts. Some work has suggested that performing dedicated movements, such as gestures, can increase the amount of usable entropy from motion sensors in the region of approximately 5--6 bits~\cite{lv2020analysis,voris2011accelerometers}. Our datasets do not cover such dedicated movements; it is possible that the reported results are an underestimation of entropy for sensors which are deliberately perturbed as an \emph{entropy-generating} action. Conversely, we do not examine \emph{entropy-reducing} threats, such as sensor spoofing~\cite{shrestha2018sensor}, hardware attacks~\cite{shepherd2021physical}, or cross-device correlation attacks~\cite{dautov2019effects}. It is concievable that these threats may reduce the entropy of sensors even further.\footnote{By way of example, we could consider a voltage-based fault injection attack that causes nondeterministic errors in the output of individual sensors, affecting the entropy of their output values.}

Throughout this work,, we report unconditioned, per-sample entropies for each sensor. This treats consecutive samples as independent and identically distributed (i.i.d.), which yields an upper bound on unpredictability because real sensor streams are time-correlated. For any additional side information, $Y$ (e.g.\ past samples), conditioning cannot increase entropy, i.e.\ $H_{\alpha}(X \, | \, Y) \leq H_{\alpha}(X)$ (see~\cite{iwamoto2013information} for additional properties). We may wish to condition $X$ on a sequence of previously observed measurements; however, efficiently quantifying $H_{\alpha}(X_t \, | \, X_{<t})$ for high-dimensional sensor streams remains non-trivial.  Similarly, we may consider the entropy of a \emph{block} of measurements over time. Under na\"{i}ve i.i.d.\ conditions, and for intuition purposes only, the entropy rate of a source using block size $N$ and sampling frequency, $f_s$, can be considered at most $f_s\, \cdot \,N $; for the average modality in PerilZIS (3.912), this na\"{i}vely yields a maximum of $\approx$39 bits/sec under a 10Hz sampling frequency. Any temporal dependence can only reduce this rate (cf.\ \S6, NIST SP 800-90B~\cite{turan2018recommendation}). A full temporal analysis warrants a dedicated study, but we posit that its effects would be to decrease the figures reported in this work, reinforcing our conclusion that mobile sensors are weak sources of unpredictability for security applications.
\section{Conclusion}
\label{sec:conc}

This paper presented the largest study to date on the entropy of mobile sensors across four public datasets (UCI-HAR, SHL, Relay, and PerilZIS) and 25 modalities. Our results expose a gap between the \emph{perceived} and \emph{actual} strength of sensor data for security: even when max- or Shannon entropies appear high, worst-case min-entropy often collapses to insecure levels. Modalities that appear strong under average-case metrics, which can mask inherent biases, are insecure under worst-case measures advocated in modern security standards. Consequently, this creates the opportunity for adversaries to guess the most probable values with minimal effort.

Our findings also challenge the common practice that model evaluation metrics (e.g.\ accuracy or EER) suffice to demonstrate that sensor-based security applications are, indeed, secure from a data-centric perspective. The vulnerability of mobile sensors to biased distributions and inter-sensor correlations significantly undermines their effectiveness as reliable sources of unpredictability. The effectiveness of countermeasures remains an open and pressing research challenge without compromising security or utility. Our hope is that the methodologies and insights presented here will encourage the security community to adopt more rigorous evaluation strategies for sensor-based techniques, paving the way for safer and more robust designs.

In future work, we consider that a dynamic analysis is important to assess the stationarity issues with sensor data, where entropy varies according to user behavior or environment. A user study could yield empirical bounds on how finely humans can \emph{intentionally} manipulate motion sensors, revealing more realistic limits to sensor-based randomness in real-world scenarios. Our binning decision was a statistical one, rather than one that reflects human usage; it is possible that real-world usage may reduce the resolution of useful sensor data, reducing entropy even further. Overall, while sensor-based data can reach limited levels of entropy under favourable conditions, the road to making such sources systematically secure  and robust is long. The key takeaway is that substantial work is needed before mobile sensors can be considered appropriate for security-critical applications.

\section*{Acknowledgments} Carlton Shepherd received funding from the UK's EPSRC (EP/Y030168/1) and would like to thank Darren Hurley-Smith for insightful conversations on the topic.

\appendix
\numberwithin{figure}{section}
\setcounter{figure}{0}
\renewcommand{\thefigure}{\Alph{section}.\arabic{figure}}
\section{Mutual Information Heatmaps}
\label{app:mi}
\begin{figure*}[t!]
    \centering
    \begin{subfigure}[t]{0.45\textwidth}
        \centering
        \includegraphics[width=\textwidth]{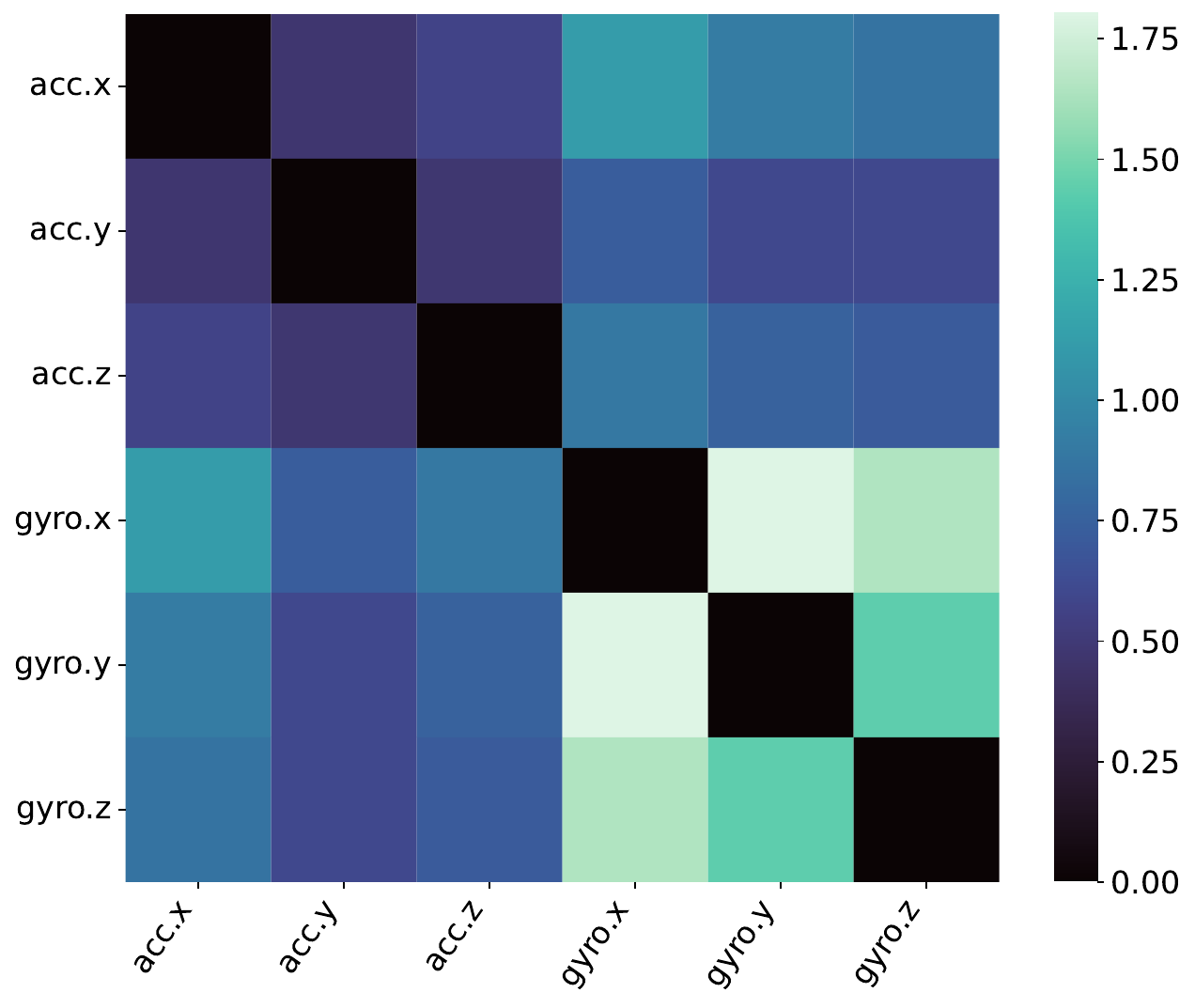}
        \caption{UCI-HAR}
        \label{subfig:umi}
    \end{subfigure}%
    \begin{subfigure}[t]{0.5\textwidth}
        \centering
        \includegraphics[width=\textwidth]{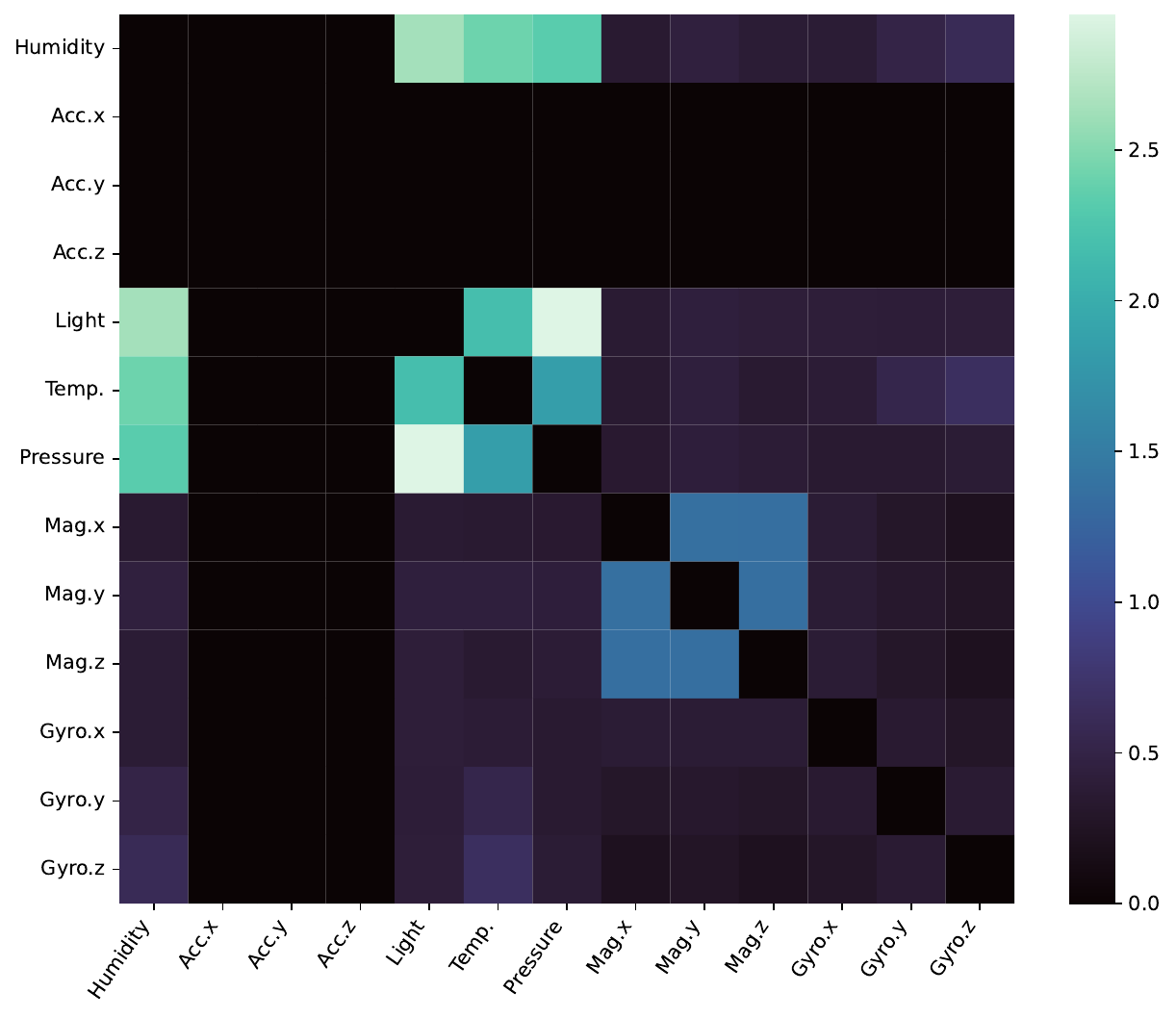}
        \caption{PerilZIS}
    \end{subfigure}%

    \begin{subfigure}[t]{0.45\textwidth}
        \centering
        \includegraphics[width=\textwidth]{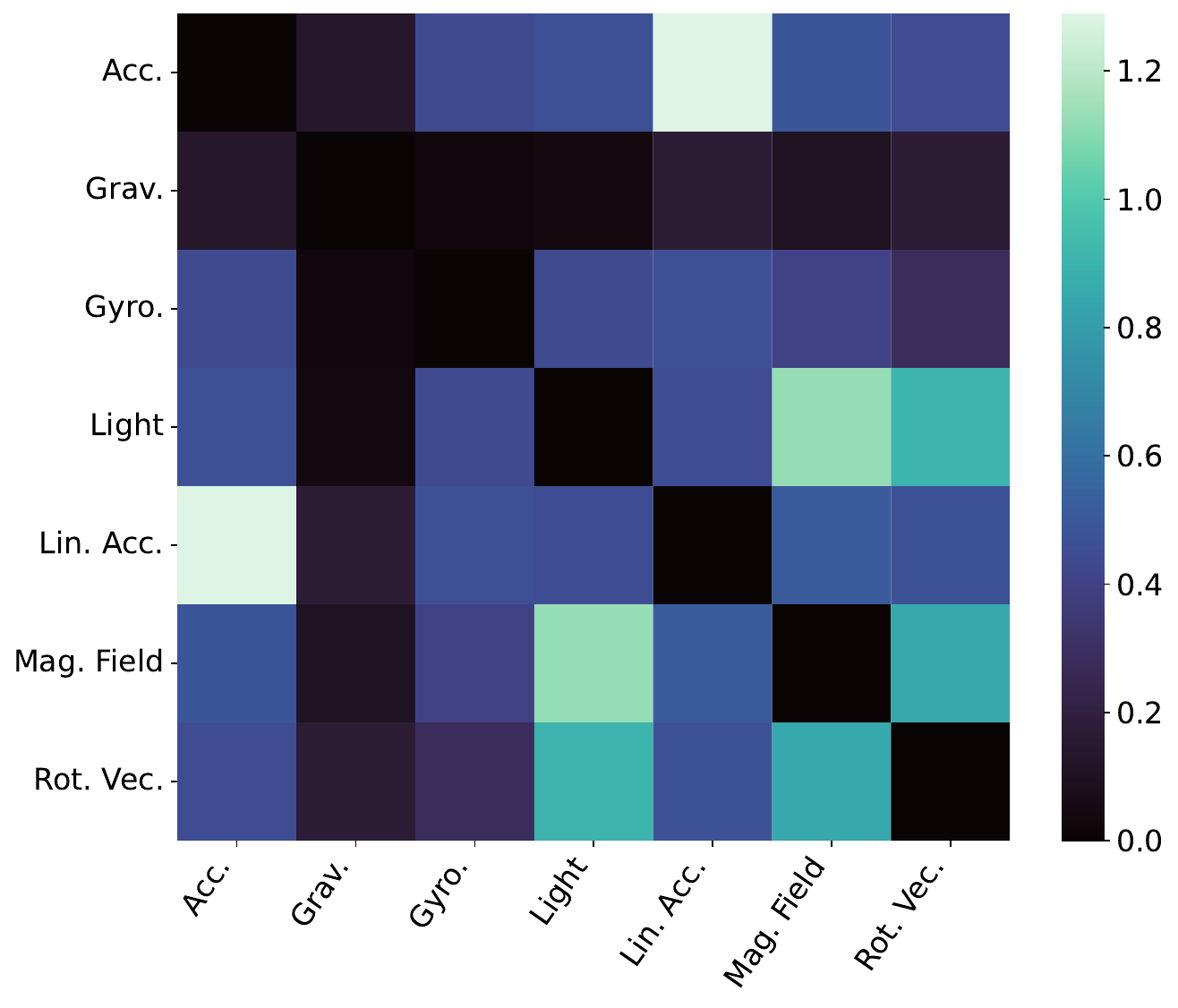}
        \caption{Relay}
    \end{subfigure}%
    \begin{subfigure}[t]{0.5\textwidth}
        \centering
        \includegraphics[width=\textwidth]{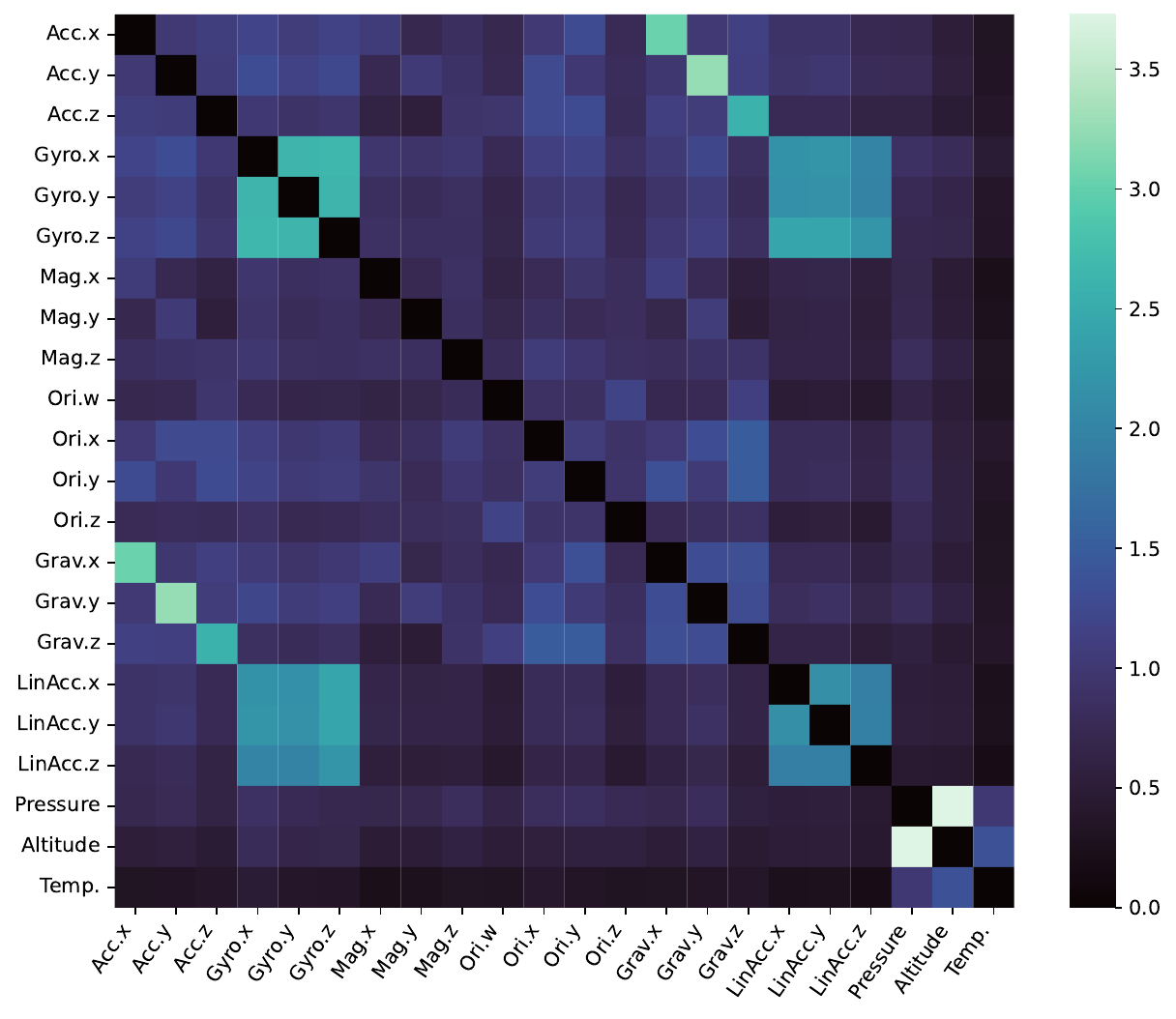}
        \caption{SHL}
    \end{subfigure}%
    \caption{Pairwise mutual information (in bits) heatmaps for each dataset.}
    \label{fig:mi-matrices}
\end{figure*}

Figure~\ref{fig:mi-matrices} shows the pairwise mutual information (in bits) between the sensors of each dataset.

{\normalsize
\bibliographystyle{elsarticle-num}
\bibliography{bibliography}

\begin{thebibliography}{10}
\expandafter\ifx\csname url\endcsname\relax
  \def\url#1{\texttt{#1}}\fi
\expandafter\ifx\csname urlprefix\endcsname\relax\def\urlprefix{URL }\fi
\expandafter\ifx\csname href\endcsname\relax
  \def\href#1#2{#2} \def\path#1{#1}\fi

\bibitem{gjoreski2018university}
H.~Gjoreski, M.~Ciliberto, L.~Wang, F.~J.~O. Morales, S.~Mekki, S.~Valentin, D.~Roggen, {University of Sussex--Huawei} locomotion and transportation dataset for multimodal analytics with mobile devices, IEEE Access (2018).

\bibitem{wang2019enabling}
L.~Wang, H.~Gjoreski, M.~Ciliberto, S.~Mekki, S.~Valentin, D.~Roggen, Enabling reproducible research in sensor-based transportation mode recognition with the {Sussex--Huawei} dataset, IEEE Access 7 (2019) 10870--10891.

\bibitem{anguita2013public}
D.~Anguita, A.~Ghio, L.~Oneto, X.~Parra, J.~L. Reyes-Ortiz, et~al., A public domain dataset for human activity recognition using smartphones, in: Esann, 2013.

\bibitem{yin2024systematic}
Y.~Yin, L.~Xie, Z.~Jiang, F.~Xiao, J.~Cao, S.~Lu, A systematic review of human activity recognition based on mobile devices: Overview, progress and trends, IEEE Communications Surveys \& Tutorials 26~(2) (2024) 890--929.

\bibitem{gurulian2017effectiveness}
I.~Gurulian, C.~Shepherd, E.~Frank, K.~Markantonakis, R.~Akram, K.~Mayes, On the effectiveness of ambient sensing for {NFC}-based proximity detection by applying relay attack data, in: 16th IEEE International Conference on Trust, Security and Privacy in Computing and Communications, Vol.~17, 2017.

\bibitem{gurulian2018good}
I.~Gurulian, K.~Markantonakis, E.~Frank, R.~N. Akram, Good vibrations: artificial ambience-based relay attack detection, in: 17th IEEE Int'l Conf. on Trust, Security and Privacy In Computing and Communications, IEEE, 2018.

\bibitem{mehrnezhad2015tap}
M.~Mehrnezhad, F.~Hao, S.~F. Shahandashti, Tap-tap and pay ({TTP}): Preventing the mafia attack in nfc payment, in: Proceedings of the 2nd International Conference on Security Standardisation Research, SSR, Springer, 2015, pp. 21--39.

\bibitem{mayrhofer2009shake}
R.~Mayrhofer, H.~Gellersen, Shake well before use: Intuitive and secure pairing of mobile devices, IEEE Transactions on Mobile Computing (2009).

\bibitem{shrestha2014drone}
B.~Shrestha, N.~Saxena, H.~T.~T. Truong, N.~Asokan, Drone to the rescue: Relay-resilient authentication using ambient multi-sensing, in: International Conference on Financial Cryptography and Data Security, Springer, 2014, pp. 349--364.

\bibitem{shrestha2018sensor}
B.~Shrestha, N.~Saxena, H.~T.~T. Truong, N.~Asokan, Sensor-based proximity detection in the face of active adversaries, IEEE Transactions on Mobile Computing 18~(2) (2018) 444--457.

\bibitem{shepherd2017applicability}
C.~Shepherd, I.~Gurulian, E.~Frank, K.~Markantonakis, R.~N. Akram, E.~Panaousis, K.~Mayes, The applicability of ambient sensors as proximity evidence for {NFC} transactions, in: IEEE Security and Privacy Workshops, IEEE, 2017, pp. 179--188.

\bibitem{shi2011senguard}
W.~Shi, J.~Yang, Y.~Jiang, F.~Yang, Y.~Xiong, Senguard: Passive user identification on smartphones using multiple sensors, in: 7th Int'l Conf. on Wirless and Mobile Computing, Networking and Communications, IEEE, 2011.

\bibitem{miettinen2014conxsense}
M.~Miettinen, S.~Heuser, W.~Kronz, A.-R. Sadeghi, N.~Asokan, Conxsense: automated context classification for context-aware access control, in: 9th ACM Symposium on Information, Computer and Communications Security, 2014.

\bibitem{li2013unobservable}
L.~Li, X.~Zhao, G.~Xue, Unobservable re-authentication for smartphones., in: Network and Distributed System Security, Citeseer, 2013.

\bibitem{li2020t2pair}
X.~Li, Q.~Zeng, L.~Luo, T.~Luo, T2pair: Secure and usable pairing for heterogeneous {IoT} devices, in: {ACM} Computer and Communications Security, 2020.

\bibitem{riva2012progressive}
O.~Riva, C.~Qin, K.~Strauss, D.~Lymberopoulos, Progressive authentication: deciding when to authenticate on mobile phones, in: 21st USENIX Security Symposium, 2012, pp. 301--316.

\bibitem{krhovjak2007sources}
J.~Krhovj{\'a}k, P.~{\v{S}}venda, V.~Maty{\'a}{\v{s}}, et~al., The sources of randomness in mobile devices, in: Proceedings of 12th Nordic Conference on Secure IT Systems, 2007.

\bibitem{xu2021key}
W.~Xu, J.~Zhang, S.~Huang, C.~Luo, W.~Li, Key generation for {Internet of Things}: A contemporary survey, ACM Computing Surveys 54~(1) (2021) 1--37.

\bibitem{abuhamad2020sensor}
M.~Abuhamad, A.~Abusnaina, D.~Nyang, D.~Mohaisen, Sensor-based continuous authentication of smartphones’ users using behavioral biometrics: A contemporary survey, IEEE Internet of Things Journal 8~(1) (2020) 65--84.

\bibitem{mayrhofer2021adversary}
R.~Mayrhofer, S.~Sigg, Adversary models for mobile device authentication, ACM Computing Surveys 54~(9) (2021) 1--35.

\bibitem{markantonakis2024using}
K.~Markantonakis, J.~A. Meister, I.~Gurulian, C.~Shepherd, R.~N. Akram, S.~A. Ghazalah, M.~Kasi, D.~Sauveron, G.~Hancke, Using ambient sensors for proximity and relay attack detection in {NFC} transactions: A reproducibility study, IEEE Access (2024).

\bibitem{truong2014comparing}
H.~T.~T. Truong, X.~Gao, B.~Shrestha, N.~Saxena, N.~Asokan, P.~Nurmi, Comparing and fusing different sensor modalities for relay attack resistance in zero-interaction authentication, in: IEEE Int'l Conf. on Pervasive Computing and Communications, IEEE, 2014.

\bibitem{nistsp80057}
E.~Barker, A.~Roginsky, Recommendation for key management: Part 1 -- general, NIST Special Publication 800-57 Part 1, Rev. 5, National Institute of Standards and Technology (May 2020).

\bibitem{turan2018recommendation}
M.~S. Turan, E.~Barker, J.~Kelsey, K.~A. McKay, M.~L. Baish, M.~Boyle, et~al., Recommendation for the entropy sources used for random bit generation, NIST Special Publication 800~(90B) (2018) 102.

\bibitem{bsi2024ais31}
{Bundesamt für Sicherheit in der Informationstechnik (BSI)}, {A Proposal for Functionality Classes for Random Number Generators (Version 3.0)}, Tech. rep., Federal Office for Information Security (BSI) (September 2024).

\bibitem{voris2011accelerometers}
J.~Voris, N.~Saxena, T.~Halevi, Accelerometers and randomness: Perfect together, in: 4th ACM Conf. on Wireless Network Security, 2011.

\bibitem{hennebert2013entropy}
C.~Hennebert, H.~Hossayni, C.~Lauradoux, Entropy harvesting from physical sensors, in: 6th ACM Security and Privacy in Wireless and Mobile Networks, 2013.

\bibitem{lv2020analysis}
N.~Lv, T.~Chen, Y.~Ma, Analysis on entropy sources based on smartphone sensors, in: 10th Int'l Conf. on Communication and Network Security, 2020.

\bibitem{halevi2012secure}
T.~Halevi, D.~Ma, N.~Saxena, T.~Xiang, Secure proximity detection for {NFC} devices based on ambient sensor data, in: 17th European Symposium on Research in Computer Security, ESORICS, Springer, 2012, pp. 379--396.

\bibitem{pan2018universense}
S.~Pan, C.~Ruiz, J.~Han, A.~Bannis, P.~Tague, H.~Y. Noh, P.~Zhang, Universense: {IoT} device pairing through heterogeneous sensing signals, in: 19th Int'l Workshop on Mobile Computing Systems and Applications, 2018.

\bibitem{patel2016continuous}
V.~M. Patel, R.~Chellappa, D.~Chandra, B.~Barbello, Continuous user authentication on mobile devices: Recent progress and remaining challenges, IEEE Signal Processing Magazine 33~(4) (2016) 49--61.

\bibitem{mekruksavanich2021deep}
S.~Mekruksavanich, A.~Jitpattanakul, Deep learning approaches for continuous authentication based on activity patterns using mobile sensing, Sensors (2021).

\bibitem{li2018sensor}
Y.~Li, H.~Hu, G.~Zhou, S.~Deng, Sensor-based continuous authentication using cost-effective kernel ridge regression, IEEE Access 6 (2018) 32554--32565.

\bibitem{micallef2015aren}
N.~Micallef, M.~Just, L.~Baillie, M.~Halvey, H.~G. Kayacik, Why aren't users using protection? investigating the usability of smartphone locking, in: 17th Int'l Conf. on Human-Computer Interaction with Mobile Devices and Services, 2015.

\bibitem{alzubaidi2016authentication}
A.~Alzubaidi, J.~Kalita, Authentication of smartphone users using behavioral biometrics, IEEE Communications Surveys \& Tutorials 18~(3) (2016) 1998--2026.

\bibitem{wu2024t2pair}
C.~Wu, X.~Li, L.~Luo, Q.~Zeng, {T2Pair}++: Secure and usable {IoT} pairing with zero information loss, arXiv preprint arXiv:2409.16530 (2024).

\bibitem{buller2016estimating}
D.~Buller, A.~Kaufer, Estimating min-entropy using probabilistic graphical models, in: Random Bit Generation Workshop, NIST, 2016.

\bibitem{suciu2011unpredictable}
A.~Suciu, D.~Lebu, K.~Marton, Unpredictable random number generator based on mobile sensors, in: IEEE 7th Int'l Conference on intelligent Computer Communication and Processing, IEEE, 2011, pp. 445--448.

\bibitem{rukhin2001nist}
A.~Rukhin, et~al., {A Statistical Test Suite for Random and Pseudorandom Number Generators for Cryptographic Applications}, Tech. Rep. SP 800-22, National Institute of Standards and Technology (April 2001).

\bibitem{wallace2016toward}
K.~Wallace, K.~Moran, E.~Novak, G.~Zhou, K.~Sun, Toward sensor-based random number generation for mobile and {IoT} devices, IEEE Internet of Things Journal 3~(6) (2016) 1189--1201.

\bibitem{hurley2020quantum}
D.~Hurley-Smith, J.~Hernandez-Castro, Quantum leap and crash: Searching and finding bias in quantum random number generators, ACM Transactions on Privacy and Security 23~(3) (2020) 1--25.

\bibitem{mai2017guessability}
G.~Mai, M.-H. Lim, P.~C. Yuen, On the guessability of binary biometric templates: A practical guessing entropy based approach, in: IEEE Int'l Joint Conf. on Biometrics, IEEE, 2017.

\bibitem{uellenbeck2013quantifying}
S.~Uellenbeck, M.~D{\"u}rmuth, C.~Wolf, T.~Holz, Quantifying the security of graphical passwords: The case of android unlock patterns, in: ACM SIGSAC Conf. on Computer and Communications Security, 2013.

\bibitem{shepherd2021physical}
C.~Shepherd, K.~Markantonakis, N.~van Heijningen, D.~Aboulkassimi, C.~Gaine, T.~Heckmann, D.~Naccache, Physical fault injection and side-channel attacks on mobile devices: A comprehensive analysis, Computers \& Security (2021).

\bibitem{dautov2019effects}
R.~Dautov, G.~R. Tsouri, Effects of passive negative correlation attack on sensors utilizing physical key extraction in indoor wireless body area networks, IEEE Sensors Letters 3~(7) (2019) 1--4.

\bibitem{Mahbub_Btas2016_UMDAA02}
U.~Mahbub, S.~Sarkar, V.~M. Patel, R.~Chellappa, Active user authentication for smartphones: A challenge data set and benchmark results, in: IEEE 8th Int'l Conf. on Biometrics Theory, Applications and Systems, 2016.

\bibitem{stragapede2023behavepassdb}
G.~Stragapede, R.~Vera-Rodriguez, R.~Tolosana, A.~Morales, {BehavePassDB}: public database for mobile behavioral biometrics and benchmark evaluation, Pattern Recognition 134 (2023) 109089.

\bibitem{acien2021becaptcha}
A.~Acien, A.~Morales, J.~Fierrez, R.~Vera-Rodriguez, O.~Delgado-Mohatar, {BeCAPTCHA}: Behavioral bot detection using touchscreen and mobile sensors benchmarked on {HuMIdb}, Engineering Applications of Artificial Intelligence (2021).

\bibitem{fomichev2019perils}
M.~Fomichev, M.~Maass, L.~Almon, A.~Molina, M.~Hollick, Perils of zero-interaction security in the {Internet of Things}, Proc. ACM on Interactive, Mobile, Wearable and Ubiquitous Technologies 3~(1) (2019) 1--38.

\bibitem{bosch_bma400}
{Bosch BMA400} accelerometer, \url{https://www.bosch-sensortec.com/products/motion-sensors/accelerometers/bma400/} (2025).

\bibitem{android_motion_sensors}
{Android Developers}, Motion sensors, \url{https://developer.android.com/develop/sensors-and-location/sensors/sensors_motion} (2025).

\bibitem{chow1968approximating}
C.~Chow, C.~Liu, Approximating discrete probability distributions with dependence trees, IEEE Trans. on Information Theory 14~(3) (1968).

\bibitem{ankan2024pgmpy}
A.~Ankan, J.~Textor, pgmpy: A {Python} toolkit for {Bayesian} networks, Journal of Machine Learning Research 25~(265) (2024) 1--8.

\bibitem{arikan2002inequality}
E.~Ar{\i}kan, An inequality on guessing and its application to sequential decoding, IEEE Transactions on Information Theory 42~(1) (2002) 99--105.

\bibitem{cachin1997entropy}
C.~Cachin, Entropy measures and unconditional security in cryptography, Ph.D. thesis, ETH Zurich (1997).

\bibitem{von195113}
J.~Von~Neumann, Various techniques used in connection with random digits, Appied Math Series 12~(36-38) (1951) 5.

\bibitem{trevisan2001extractors}
L.~Trevisan, et~al., Extractors and pseudorandom generators, Journal of the ACM 48~(4) (2001) 860--879.

\bibitem{barak2003true}
B.~Barak, R.~Shaltiel, E.~Tromer, True random number generators secure in a changing environment, in: Cryptographic Hardware and Embedded Systems, Springer, 2003, pp. 166--180.

\bibitem{liu2025post}
L.~Liu, J.~Yang, M.~Wu, J.~Liu, W.~Huang, Y.~Li, B.~Xu, A post-processing method for quantum random number generator based on zero-phase component analysis whitening, Entropy 27~(1) (2025) 68.

\bibitem{vadhan2012pseudorandomness}
S.~P. Vadhan, et~al., Pseudorandomness, Foundations and Trends in Theoretical Computer Science 7~(1--3) (2012) 1--336.

\bibitem{iwamoto2013information}
M.~Iwamoto, J.~Shikata, Information theoretic security for encryption based on conditional {R{\'e}nyi} entropies, in: International Conference on Information Theoretic Security, Springer, 2013, pp. 103--121.

\end{thebibliography}
}
\end{document}